\documentclass{article}

\PassOptionsToPackage{numbers, compress}{natbib}

\usepackage[final,main]{neurips_2026}

\usepackage[utf8]{inputenc}
\usepackage[T1]{fontenc}
\usepackage{hyperref}
\usepackage{url}
\usepackage{booktabs}
\usepackage{tabularx}
\usepackage{amsfonts}
\usepackage{amsmath,amssymb,amsthm,mathtools,bm}
\usepackage{nicefrac}
\usepackage{microtype}
\usepackage{xcolor}
\usepackage{graphicx}
\usepackage{subcaption}

\newtheorem{theorem}{Theorem}

\newtheorem{lemma}{Lemma}
\newtheorem{corollary}{Corollary}

\newcommand{\R}{\mathbb{R}}

\newcommand{\E}{\mathbb{E}}
\newcommand{\diag}{\mathrm{diag}}
\newcommand{\tr}{\mathrm{tr}}
\newcommand{\argmax}{\mathop{\mathrm{arg\,max}}}

\title{PrismQuant: Rate--Distortion-Optimal Vector Quantization for Gaussian-Mixture Sources}

\author{%
  Bumsu Park \\
  Department of Electrical Engineering\\
  POSTECH \\
  Pohang, South Korea \\
  \texttt{bumsupark@postech.ac.kr} \\
  \And
  Chanho Park \\
  Department of Electrical Engineering\\
  POSTECH \\
  Pohang, South Korea \\
  \texttt{chanho26@postech.ac.kr} \\
  \AND
  Youngmok Park \\
  Department of Electrical Engineering\\
  POSTECH \\
  Pohang, South Korea \\
  \texttt{ympark1999@postech.ac.kr} \\
  \And
  Namyoon Lee\thanks{Corresponding author.} \\
  Department of Electrical Engineering\\
  POSTECH \\
  Pohang, South Korea \\
  \texttt{nylee@postech.ac.kr} \\
}

\begin{document}

\maketitle

\begin{abstract}
For a Gaussian source under mean-squared error (MSE), classical transform coding is rate--distortion (RD) optimal: the Karhunen--Lo\`eve transform (KLT) diagonalizes the covariance, reverse waterfilling allocates the bits, and scalar quantization closes the loop. This elegant story breaks down for \emph{multimodal} sources, where no single covariance can capture heterogeneous local geometries and the RD function loses its closed form. We revisit this problem through Gaussian-mixture sources and develop a constructive RD theory for them. Our key finding is that \emph{mixture structure costs only a component label}. Conditioned on the active mixture component, each branch is Gaussian; the challenge is allocating bits across heterogeneous branches. We prove that the genie-aided conditional RD function is governed by a \emph{single global reverse-waterfilling level} shared across all components and eigenmodes. Building on this result, we introduce \textsc{PrismQuant}, which transmits the component label losslessly and encodes the residual using the component-matched KLT followed by scalar quantization, achieving a rate within $H(C)/n$ bits per source dimension of the converse with a vanishing asymptotic gap. We further develop a practical implementation based on EM-driven Gaussian-mixture learning, component-adaptive KLTs, and entropy-constrained scalar quantization (ECSQ). Experiments on synthetic Gaussian mixtures show that \textsc{PrismQuant} closely approaches the theoretical RD bound, while experiments on real-world channel-state-information (CSI) data demonstrate competitive or superior performance compared with transformer-based learned codecs at more than one order of magnitude smaller model size.

\end{abstract}

\section{Introduction}
\label{sec:intro}

Shannon's rate--distortion (RD) theorem specifies the minimum rate required to represent a source at a prescribed distortion \citep{Shannon59,Berger71,CoverThomas06}. For high-dimensional sources, however, the theorem is largely existential: under mean-squared error (MSE), the Gaussian source remains the only mainstream setting aadmitting an \emph{explicit, constructive} RD characterization via the Karhunen--Lo\`eve transform (KLT) and reverse waterfilling \citep{Berger71,Goyal01}. This classical framework breaks down for \emph{multimodal} sources, where a single covariance structure cannot adequately capture heterogeneous local geometries, and the exact RD function has remained open since Berger's 1971 monograph \citep{Berger71}.

\paragraph{Setting.}
We focus on the most analytically tractable and operationally meaningful class of multimodal sources: \emph{Gaussian mixtures}. We use $X\in\R^n$ to denote the random source vector and $x\in\R^n$ a generic realization. The source distribution is
\begin{align} \label{eq:mixture}
p_X(x) = \sum_{c=1}^{K}\pi_c\,\mathcal N(x;\,\mu_c, R_c),
\end{align}
where $C\in[K]$ is a discrete mixture component label over $K$ mixture components with prior $\{\pi_c\}_{c=1}^K$. Conditioned on $C = c$, the source follows $X\sim\mathcal N(\mu_c, R_c)$ with mean $\mu_c\in\R^n$ and covariance $R_c\in\R^{n\times n}$. The Gaussian-mixture model (GMM) class is \emph{far less restrictive than it first appears}: Gaussian mixtures are universal density approximators \citep{Bishop06}, meaning that any continuous integrable density can be approximated well in $L^1$ distance by a sufficiently rich mixture (Lemma~\ref{lem:univ}). Consequently, a constructive RD theory for \eqref{eq:mixture} naturally extends to \emph{arbitrary} continuous multimodal sources by combining expectation--maximization (EM) mixture learning with the proposed coding framework.

\paragraph{The CSI compression as a motivating application.} 
Compression of high-dimensional multimodal sources is a fundamental challenge in modern data-driven systems. A particularly important example arises in frequency-division-duplex (FDD) massive multiple-input multiple-output (MIMO) orthogonal-frequency-division-multiplexing (OFDM) wireless systems, where accurate channel-state information (CSI) feedback is critical for reliable communication \citep{Marzetta10MM,Love08LimitedFB}. Since CSI is inherently high-dimensional due to large antenna arrays and many subcarriers, compression is unavoidable under strict bandwidth constraints. While CSI under fixed propagation conditions is well approximated by a correlated Gaussian model \citep{Heath16MIMO}, realistic CSI aggregated across users and environments exhibits strongly multimodal structure, naturally motivating Gaussian-mixture modeling \citep{Liu12COST2100}. In this regime, single-covariance KLT coding becomes fundamentally mismatched, whereas modern learned CSI codecs \citep{Wen18CsiNet,Park25ICCNTC} often rely on large black-box neural architectures. \textsc{PrismQuant} addresses this gap through an explicit constructive RD theory together with a low-complexity and interpretable coding framework.

% Compression of high-dimensional multimodal sources is a central challenge in modern data-driven systems. A particularly important instance arises in frequency-division-duplex (FDD) massive multiple-input multiple-output (MIMO) orthogonal-frequency-division-multiplexing (OFDM) wireless systems, where accurate channel-state information (CSI) feedback is critical for reliable communication \citep{Marzetta10MM,Love08LimitedFB}. Without accurate CSI shared between the transmitter and receiver, communication performance can degrade significantly. Moreover, CSI is inherently high-dimensional, as modern wireless channels are represented through large antenna arrays and many subcarriers, making compression unavoidable under strict bandwidth and latency constraints. Under fixed propagation conditions, CSI is well approximated by a correlated proper complex Gaussian \citep{Heath16MIMO}. In realistic deployments, however, CSI aggregated across users, locations, and blockage conditions exhibits strongly multimodal structure, for which Gaussian mixtures provide a natural statistical model \citep{Liu12COST2100}. In this regime, single-covariance KLT coding becomes fundamentally mismatched, while modern learned CSI codecs \citep{Wen18CsiNet,Park25ICCNTC} often rely on large black-box neural architectures. \textsc{PrismQuant} addresses this gap through an explicit constructive RD theory together with a low-complexity, interpretable, and data-driven coding framework.

\paragraph{Contributions.}

\textbf{(i) First explicit constructive RD characterization for Gaussian-mixture sources.}
We extend the explicit, constructive RD theory, previously known only for single-covariance Gaussian sources, to general Gaussian mixtures via a converse-achievability \emph{sandwich}. The resulting converse--achievability gap is exactly the amortized label entropy $H(C)/n$ and vanishes asymptotically whenever the mixture order satisfies $K=2^{o(n)}$. By Gaussian-mixture universality, the result further lifts to arbitrary multimodal sources through EM-based mixture fitting. Conditional source coding \citep{Berger71,CoverThomas06} provides a general theoretical framework, but obtaining an explicit RD characterization together with a constructive code for multimodal sources has remained elusive. \emph{Why has this been hard?} The difficulty stems from the fact that the unconditional Gaussian-mixture density admits no tractable entropy characterization; even sharp bounds and closed-form approximations for mixture entropy are themselves nontrivial \citep{LeeEntropyMix2026}. Our key idea is to augment the source with the latent component label, converting the non-Gaussian entropy into a sum of Gaussian entropies plus an exact label cost $H(C)$. This yields a separable convex optimization whose Karush-Kuhn-Tucker (KKT) conditions induce a single global reverse-waterfilling level shared across all mixture components.

% \textbf{(i) First explicit RD characterization for Gaussian-mixture
% sources, with universality.} We extend the explicit, constructive RD
% theory---known until now only for single-covariance Gaussian
% sources---to general Gaussian mixtures via a converse--achievability
% \emph{sandwich} whose gap is exactly the amortized label entropy
% $H(C)/n$ and vanishes whenever $\log_2(K_n)/n\to 0$ as $n\to\infty$,
% which is a strictly weaker condition than fixed $K$ (it admits any
% sub-exponential growth $K_n=2^{o(n)}$). \chanho{$K_n$    .                 $H(C)$  $K_n$               ?              ?} By Gaussian-mixture universality the result lifts, by composition, to \emph{arbitrary} multimodal
% sources via EM-based fitting. Conditional source-coding theory \citep{Berger71,CoverThomas06} provides a general framework, but an explicit RD characterization for multimodal sources together with a constructive code attaining the bound has remained elusive. 
% \emph{Why has this been hard?} The
% unconditional density $p_X$ of a Gaussian mixture is non-Gaussian and
% non-log-concave; its differential entropy admits no closed form, so
% plug-in attempts to replicate the single-Gaussian KLT argument fail.
% Our route is to introduce the latent component label as augmented side
% information, which converts the non-Gaussian entropy into a sum of
% Gaussian entropies plus an exact label cost $H(C)$; then the global
% water level emerges from the resulting separable convex
% Karush--Kuhn--Tucker (KKT) system.

\textbf{(ii) PrismQuant codec.} We propose \textsc{PrismQuant}, a practical two-stage lossless--lossy coding framework for Gaussian-mixture sources that attains the achievability bound under correct label detection. The codec first learns the underlying Gaussian-mixture distribution via expectation--maximization (EM), estimates the component label through maximum-a-posteriori (MAP) inference, and transmits the label losslessly. Conditioned on the detected component, \textsc{PrismQuant} then applies the component-matched KLT followed by entropy-constrained scalar quantization (ECSQ) and entropy coding across the resulting eigenmodes. The entire pipeline admits closed-form operations at every stage and yields a fully practical bitstream-level implementation with low complexity and strong interpretability.

% \textbf{(ii) PrismQuant codec.} We propose \textsc{PrismQuant}, a
% two-stage lossless--lossy code for Gaussian mixtures that attains the
% achievability bound under correct label detection. The codec sends the
% mixture label losslessly and then applies the component-matched KLT
% plus scalar quantization with entropy coding.

\textbf{(iii) One global reverse-waterfilling level.} 
Both the converse and the achievability are governed by a \emph{single global} reverse-waterfilling level $\mu^\star$ shared across the eigenmodes of \emph{all} mixture components. The Gaussian-mixture RD problem therefore collapses onto the same one-dial allocation structure as classical single-covariance Gaussian transform coding (TC), despite the presence of heterogeneous eigenbases. This is fundamentally different from prior classified TC methods such as WUTC \citep{EffrosChou95}, which use independent per-component water levels $\mu_c^\star$ and therefore cannot perform globally optimal rate allocation with respect to the full mixture prior. In contrast, the shared water level in \textsc{PrismQuant} induces more practical cross-component eigenmode pruning and yields the genie-aided conditional RD characterization up to the amortized label cost $H(C)/n$.

% \textbf{(iii) One global reverse-waterfilling level.} Both the converse
% and the achievability are governed by a \emph{single} global
% reverse-waterfilling level $\mu$ applied across the eigenmodes of
% \emph{all} mixture components. The mixture problem therefore collapses
% onto the same one-dial structure as the classical Gaussian KLT
% allocation, with cross-component pruning across heterogeneous
% eigenbases.

\textbf{(iv) Validation on synthetic mixtures and real-world CSI data.} 
On synthetic Gaussian-mixture sources, we sweep the mixture order $K$ at fixed source dimension $n$ and show that \textsc{PrismQuant} with MAP-estimated component labels closely tracks both the genie-aided oracle and the theoretical RD bound across varying $K$, while single-covariance TC exhibits an increasingly severe mismatch gap as multimodality grows. On the real-world DeepMIMO dataset \cite{alkhateeb2019deepmimo}, \textsc{PrismQuant} substantially improves over classical TC and achieves competitive reconstruction performance against learned CSI compression schemes including CsiNet \citep{Wen18CsiNet} and Swin-NTC \citep{Park25ICCNTC}, while requiring one to two orders of magnitude smaller model size.

% \textbf{(iv) Validation on synthetic mixtures and a real CSI
% dataset.} On synthetic Gaussian-mixture sources, we sweep the mixture
% order $K$ at fixed source dimension $n$ and show that
% \textsc{PrismQuant} with maximum-a-posteriori (MAP) labels tracks the
% genie-aided oracle and the theoretical upper bound across $K$, while
% single-covariance transform coding (TC) widens its mismatch gap as
% $K$ grows. On the real-world COST2100 massive
% MIMO CSI dataset, \textsc{PrismQuant} substantially improves over
% single-covariance KLT coding and is competitive with the
% transformer-based Swin-NTC and CsiNet baselines at one to two orders
% of magnitude lower model size.

\section{Related Works}
\label{sec:related}

We organize prior work into four lines and clarify, line by line, how \textsc{PrismQuant} differs from each. A point-by-point architectural comparison is deferred to Appendix~\ref{app:tc_wutc_pq}.

\paragraph{Conditional and side-information source coding.}
Conditional rate--distortion theory \citep{Gray72CondRD,Berger71}, Wyner--Ziv coding \citep{WynerZiv76}, and Heegard--Berger coding \citep{HeegardBerger85} provide general theoretical frameworks for structured compression problems. These frameworks characterize the limits but do not yield an explicit closed-form RD function for the multimodal source classes targeted in this paper; \textsc{PrismQuant} can be read as a constructive specialization that supplies such a closed form for Gaussian mixtures.

\paragraph{Classified vector quantization and weighted universal transform coding (WUTC).}
The closest architectural ancestor of \textsc{PrismQuant} is the line of classified VQ \citep{RiskinGray91} and Weighted Universal Transform Coding \citep{EffrosChouGray94,EffrosChou95}. These schemes also transmit a latent component label losslessly and then encode the residual using a component-matched Karhunen--Lo\`eve transform (KLT) plus scalar quantization. There are two structural differences. \emph{First}, the branch-selection rule: WUTC selects the active class at the encoder by an exhaustive rate--distortion search that encodes the block under every candidate transform and keeps the best, whereas \textsc{PrismQuant} uses a \emph{Bayes-optimal MAP classifier} $\widehat C(x)=\arg\max_c \pi_c\mathcal N(x;\mu_c,\Sigma_c)$ derived from the fitted Gaussian-mixture source model. This replaces $K$ encode-and-measure evaluations per block with a single Mahalanobis-distance comparison, and identifies the branch entropy with the latent-label entropy $H(C)$. \emph{Second}, the rate-allocation rule: WUTC uses an \emph{independent} reverse-waterfilling level $\mu_c^\star$ per component, whereas \textsc{PrismQuant} uses a \emph{single global} water level $\mu^\star$ shared across all components and eigenmodes (Theorem~\ref{thm:water}). Together, these two changes convert a heuristic multi-codebook codec without a matching converse into one that attains the genie-aided RD function up to the amortized label cost $H(C)/n$ (Theorem~\ref{thm:ach}). Closely related Gaussian-mixture VQ schemes \citep{LindeBuzoGray80,HedelinSkoglund00} share the multi-codebook spirit but likewise lack both a probabilistic selection rule and a converse--achievability sandwich, and the classical single-covariance Gaussian transform coder \citep{Berger71,Goyal01} is recovered as the special case $K=1$.

\paragraph{Adaptive multiple transform (AMT) in standardized video codecs.}
A standardized, hand-designed instance of the library-of-transforms idea is the \emph{adaptive multiple transform} (AMT)---also called Multiple Transform Selection (MTS)---adopted in HEVC and VVC. The concept was first introduced by Han, Saxena, and Rose \citep{HanSaxenaRose10ASTC,HanSaxenaMelkoteRose12AMT} as a joint optimization of spatial prediction and block transform; it was extended into a mode-dependent DCT/DST coding scheme by Saxena and Fernandes \citep{SaxenaFernandes13AMT}, generalized as enhanced multiple transform by Zhao et al.\ \citep{ZhaoEMT16}, and standardized in VVC \citep{BrossVVC21}. AMT also signals a branch index, but its library consists of a small number of \emph{fixed trigonometric transforms} (DCT-II together with DST-VII and DCT-VIII variants), the branch is chosen by an encoder-side rate--distortion search, and the analysis is purely operational. \textsc{PrismQuant} differs from AMT in three respects: (i) its library is derived from a fitted Gaussian-mixture source model rather than hand-designed; (ii) the branch is selected by a Bayes-optimal MAP classifier rather than by an RD search; and (iii) the bit allocation is summarized by a single global water level with a matching closed-form converse, instead of by per-block quantization parameters without an information-theoretic gap analysis.

\paragraph{Neural compression and RD estimators.}
Modern learned compressors specialize expressive neural architectures to specific source classes (e.g., image \citep{Balle18Hyperprior,cheng2020learned} and CSI \citep{Wen18CsiNet,Park25ICCNTC}). Their RD behavior is characterized empirically through training objectives and benchmark curves rather than through explicit coding laws, and plug-in or neural RD estimators \citep{10124059} likewise provide numerical bounds rather than constructive codes. Source-agnostic vector quantization such as TurboQuant \citep{Zandieh2025TurboQuant} is data-oblivious by design and aims at worst-case rather than source-adaptive performance. \textsc{PrismQuant} occupies the complementary corner of the design space: source-adaptive but driven by a probabilistic source model rather than by a learned black box, with an EM-driven Gaussian-mixture fit \citep{Bishop06,Dempster77}, MAP-based component-label inference, component-matched KLTs, and entropy-constrained scalar quantization (ECSQ).

\section{Gaussian-Mixture Source and Universality}
\label{sec:model}

\paragraph{Problem setup.}
We consider an independent and identically distributed (i.i.d.) random vector source $\{X_t\}_{t\ge 1}$ with $X_t\in\R^n$ drawn from the real Gaussian mixture \eqref{eq:mixture};
$x_t\in\R^n$ denotes a generic realization. A blocklength-$m$, rate-$R$ source code is a pair of mappings $f_m:(\R^n)^m\to\{1,\ldots,2^{mnR}\}$ and $g_m:\{1,\ldots,2^{mnR}\}\to(\R^n)^m$, where $R$ is in bits per source dimension. The fidelity criterion is normalized MSE (NMSE) $D=\tfrac{1}{mn}\sum_{t=1}^{m}\E\|X_t-\hat X_t\|_2^2$. The operational RD function $R^\star(D)$ is the infimum of all achievable rates at distortion $D$ \citep{Shannon59,Berger71,CoverThomas06}.

\paragraph{Why Gaussian mixtures?}
Many real-world sources are well modeled as locally Gaussian within a homogeneous regime, but globally non-stationary across regimes (e.g., channel responses across propagation geometries \cite{wen2015channel, song2025downlink}, image patches across textures \cite{nguyen2005study, zoran2011learning}, semantic features across classes \cite{cheng2020learned, falck2021multi}, and speech or acoustic features across speakers \cite{povey2010subspace}). A Gaussian mixture replaces the (intractable) integral over the latent regime by a finite collection of representative covariances, each conditionally Gaussian. By Lemma~\ref{lem:univ} below, the model is universal: any continuous integrable density admits an arbitrarily accurate Gaussian-mixture surrogate.

\begin{lemma}[Universality of Gaussian mixtures \citep{Bishop06}]
\label{lem:univ}
For any continuous integrable density $p$ on $\R^d$ and any $\varepsilon>0$, there exist $K$, weights $\{\pi_c\}_{c=1}^K$, means $\{\mu_c\}_{c=1}^K$, and covariances $\{\Sigma_c\}_{c=1}^K$ such that $p_K(x)=\sum_c \pi_c \mathcal N(x;\mu_c,\Sigma_c)$ satisfies $\int |p-p_K|\,dx<\varepsilon$.
\end{lemma}

Composing Lemma~\ref{lem:univ} with \textsc{PrismQuant} gives a constructive codec for an \emph{arbitrary} continuous multimodal source: learn the parameters of Gaussian mixture distribution by EM and feed it to the encoder. The explicit RD theory below is stated for the finite mixture
\eqref{eq:mixture}, but its operational reach is the entire class of continuous multimodal densities.

\begin{lemma}[Entropy decomposition]
\label{lem:entropy}
Let $C\in[K]$ have prior $\{\pi_c\}_{c=1}^K$ and let $\left. X \right| \{C=c\}\sim \mathcal N(\mu_c,R_c)$. Then $h(X)=h(\left.X\right| C)+I(X;C)$ and
\begin{align}
\label{eq:entropy_sandwich}
h(\left.X\right| C) \le h(X) \le h( \left. X\right| C)+H(C),\quad h(\left.X\right| C)=\frac{1}{2}\sum_{c=1}^{K}\pi_c\log_2\bigl[(2\pi e)^n\det R_c\bigr],
\end{align}
where the conditional differential entropy depends on $\{R_c\}_{c=1}^K$ only (component means $\{\mu_c\}_{c=1}^K$ contribute zero entropy because differential entropy is translation-invariant).
\end{lemma}

The lower endpoint $h(\left.X\right| C)$ is the genie-aided benchmark in which the active component is known for free; the upper endpoint corresponds to (1)~\emph{lossless label coding} and (2)~\emph{component-matched lossy transform coding}. This is the organizing principle of \textsc{PrismQuant}: the discrete, multimodal part of the source is absorbed into a low-rate label stream, leaving a textbook Gaussian transform-coding problem within the selected branch. The non-oracle price is exactly the normalized label cost $H(C)/n$. A complementary view of the same decomposition, including sharp bounds and closed-form approximations of the unconditional mixture entropy $h(X)$ in terms of pairwise component overlaps, is developed in \citet{LeeEntropyMix2026}, providing a route from raw mixture parameters $\{\pi_c,\mu_c,R_c\}_{c=1}^K$ to a closed-form estimate of every information measure used in our analysis.

\begin{figure}[t]
    \centering
    \includegraphics[width=0.8\textwidth]{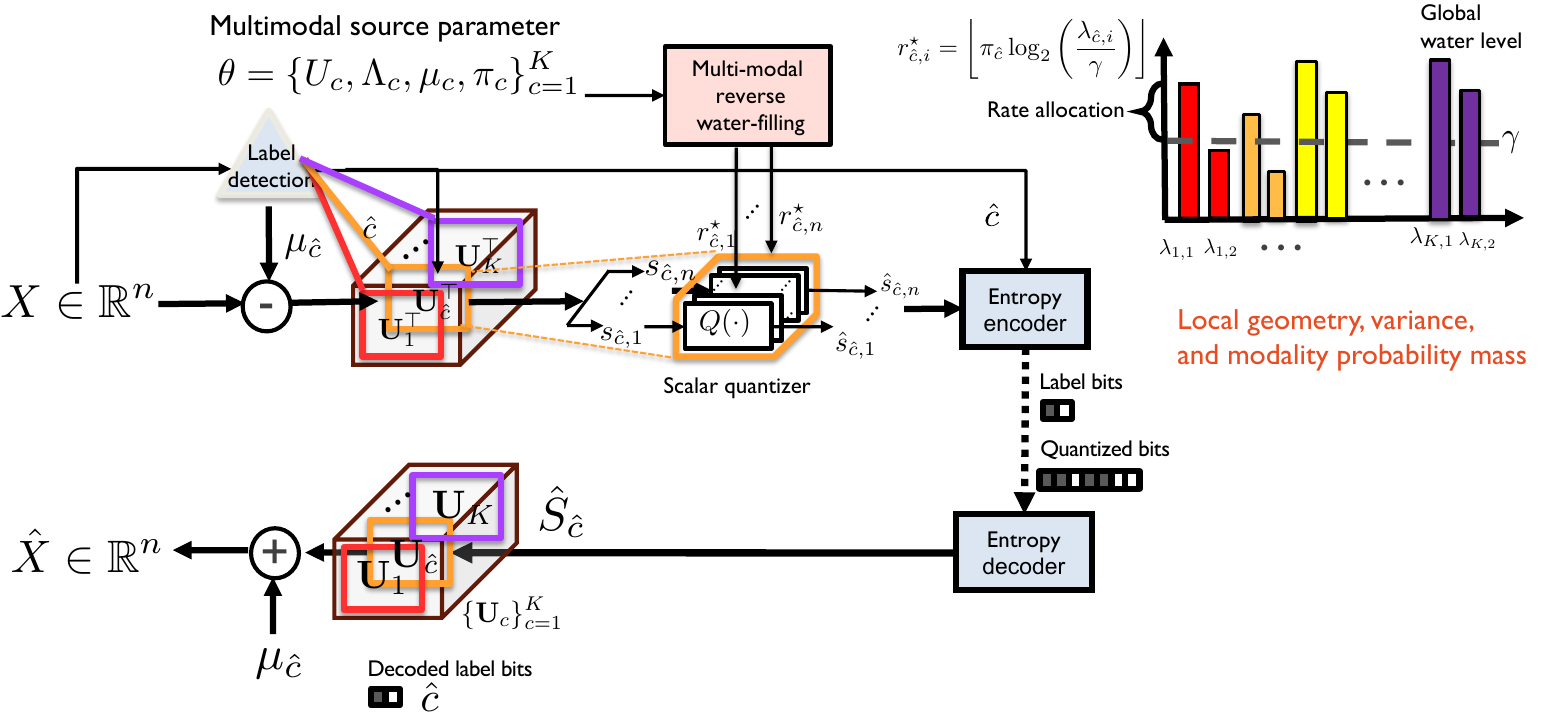}
    \caption{Encoder and decoder of \textsc{PrismQuant}. The encoder selects the most likely mixture component, transmits its label losslessly, applies the component-matched KLT, and entropy-codes the active transform coefficients. The decoder inverts the pipeline through the shared dictionary $\theta=\{(\pi_c,\mu_c,R_c)\}_{c=1}^{K}$.} 
    \label{fig:pipeline}
\end{figure}

\section{The PrismQuant Codec}
\label{sec:prismquant}

\textsc{PrismQuant} is a practical hybrid lossless--lossy coding framework for the Gaussian-mixture source \eqref{eq:mixture}. The mixture parameters $\theta=\{(\pi_c,\mu_c,R_c)\}_{c=1}^K$ are learned offline (e.g., EM on training samples) and stored as a shared probabilistic dictionary at both the encoder and decoder; only the component label and the coded transform coefficients are transmitted online. The entire pipeline admits closed-form operations at every stage and yields a fully practical bitstream-level implementation.

\paragraph{Encoder.}
Let $R_{\rm tot}$ denote the total budget in bits/dim. 
(1) \emph{Component label estimation:} the encoder selects the most likely mixture component using MAP estimation, i.e., $\hat c(x)=\argmax_{c\in[K]} \pi_c\,\mathcal N(x;\mu_c,R_c).$
(2) \emph{Lossless label coding:} the estimated label $\hat c$ is entropy coded; the normalized label rate is $R_{\rm lbl}=H(C)/n$. 
(3) \emph{Mean centering and component KLT:} the encoder forms the centered residual $s = U_{\hat c}^{\top}(x-\mu_{\hat c})$, where $R_{\hat c} = U_{\hat c}\Lambda_{\hat c}U_{\hat c}^{\top}$ is the spectral decomposition. The transform coefficients are independent and $s_i\sim \mathcal N(0,\lambda_{\hat c,i}), \, \forall i \in [n]$. 
(4) \emph{Global reverse-waterfilling:} given the residual budget $R_{\rm q}=R_{\rm tot}-R_{\rm lbl}$, a single water level $\mu\ge 0$ allocates per-eigenmode rate and distortion across \emph{all} components by
\begin{align}
\label{eq:water_alloc}
d_{c,i}^\star(\mu) = \min\{\lambda_{c,i},\,\mu\}, \qquad r_{c,i}^\star(\mu) = \tfrac{1}{2}\bigl[\log_2(\lambda_{c,i}/\mu)\bigr]_+,
\end{align}
chosen so that $R_{\rm q}=\tfrac{1}{n}\sum_c\pi_c\sum_i r_{c,i}^\star(\mu)$ (see Theorem~\ref{thm:water}). 
(5) \emph{Scalar quantization and entropy coding:} active coefficients ($\lambda_{\hat c,i}>\mu$) are quantized at the allocated rate using ECSQ \cite{GershoGray92}, while inactive coefficients are set to zero. 
(6) \emph{Reconstruction:} the decoder reconstructs $\hat x = \mu_{\hat c} + U_{\hat c}\hat s$. The overall encoding and decoding pipeline of \textsc{PrismQuant} is illustrated in Figure~\ref{fig:pipeline}.

\paragraph{Cross-component pruning.}
Because the same global water level $\mu$ thresholds the eigenvalues of \emph{every} component, eigenmodes satisfying $\lambda_{c,i}\le\mu$ need not be stored, transmitted, or processed during runtime. Consequently, \textsc{PrismQuant} retains only the active eigenmodes that receive nonzero rate allocation under reverse waterfilling, significantly reducing both the effective model footprint and the coding complexity. Classical Gaussian reverse waterfilling performs pruning only within a single fixed eigenbasis; in contrast, the multimodal setting induces pruning jointly across heterogeneous component-dependent eigenbases under a \emph{single} global allocation parameter. The resulting runtime memory and per-vector encode/decode complexity therefore scale as $\Theta(nL_c(\mu))$, where $L_c(\mu)=|\{i:\lambda_{c,i}>\mu\}|$ denotes the number of active eigenmodes in component $c$.

% \paragraph{Cross-component pruning.}
% Because the same $\mu$ thresholds the eigenvalues of \emph{every} component, eigenmodes for which $\lambda_{c,i}\le\mu$ in component $c$ need not be stored or touched at runtime. Standard Gaussian reverse waterfilling prunes eigenmodes within a single fixed eigenbasis; the multimodal extension prunes eigenmodes across heterogeneous eigenbases under \emph{one} global dial. The runtime model footprint and the per-vector encode/decode cost both scale as $\Theta(n L_c(\mu))$ with $L_c(\mu)=|\{i:\lambda_{c,i}>\mu\}|$.

\begin{figure}[t]
\centering
\setlength{\belowcaptionskip}{0pt}   % caption         
    
\includegraphics[width=0.7\textwidth]{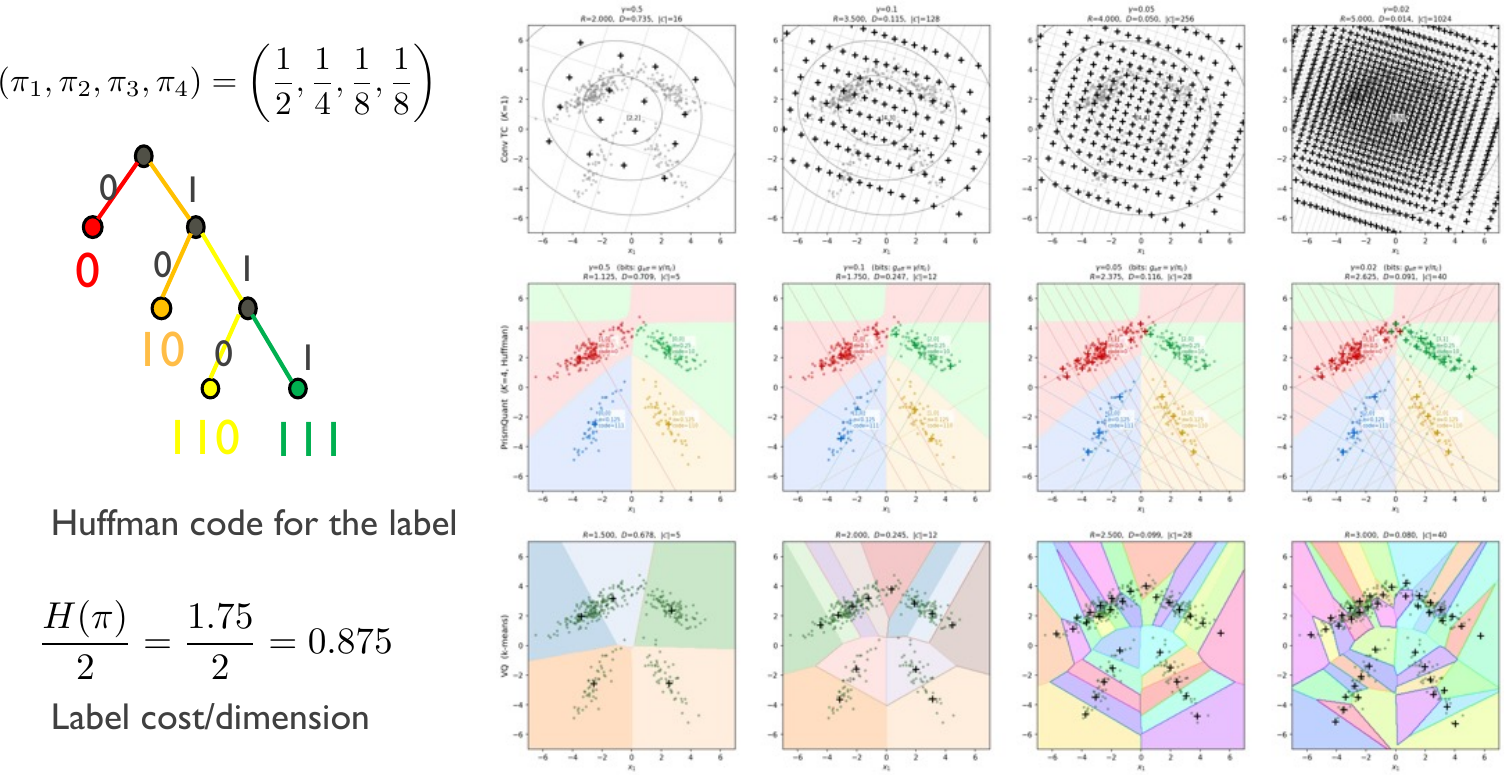}
\caption{\textbf{Left:} Huffman code for $C$ with prior
$(\tfrac12,\tfrac14,\tfrac18,\tfrac18)$, $H(\pi)=1.75$ bits/symbol.
\textbf{Right:} TC (top), \textsc{PrismQuant} (middle), and $k$-means
VQ (bottom) on a four-component mixture.}
\label{fig:label_huffman_vq}
\end{figure}

\paragraph{Label coding and TC vs.\ \textsc{PrismQuant} vs.\ VQ.}
The lossless label rate $R_{\rm lbl}=H(C)/n$ is operationally realized by a Huffman code over $C$
(Figure~\ref{fig:label_huffman_vq}, left): for the dyadic prior $(\tfrac12,\tfrac14,\tfrac18,\tfrac18)$ the expected codeword length equals $H(\pi)=1.75$ bits, attaining the entropy exactly; for general priors the gap is at most $1/n$ bits/dim with length-$n$ blocking. The right panel compares the three encoders geometrically on a four-component mixture: a single global TC (top) is mixture-oblivious; \textsc{PrismQuant} (middle) sends the label and then runs a component-matched KLT plus scalar quantization, so the per-component grids align with the local covariance ellipses; an unstructured $k$-means VQ \citep{LindeBuzoGray80} (bottom) recovers similar Voronoi geometry but pays the exponential codebook $|\mathcal C|=2^{nR}$ and substantial learning cost with complexity $O(n |\mathcal{C}|)$. Across rates (increasing from left to right), \textsc{PrismQuant} progressively resolves each mixture component while reusing one shared dictionary $\theta=\{(\pi_c,\mu_c,U_c,\Lambda_c)\}_{c=1}^{K}$.

\paragraph{Offline learning.}
The mixture is learned offline by complex EM on a representative training set $\{x_t\}_{t=1}^{N_{\rm tr}}$: the E-step computes posterior responsibilities and the M-step updates $\pi_c$ and $R_c$ via weighted second moments. The dictionary is then frozen for online use; only the label and the coded coefficients are transmitted. By Lemma~\ref{lem:univ}, the same machinery applies (with sufficient $K$) to any continuous multimodal source; the user only changes the EM data.

\section{Rate--Distortion Characterization}
\label{sec:rd}

We now derive an explicit converse--achievability sandwich for the Gaussian-mixture source \eqref{eq:mixture}.

\subsection{Conditional Gaussian benchmark and the global water level}
For each component, the classical Gaussian RD function \citep{Berger71} gives $R_c(D_c)=\tfrac{1}{n}\sum_i \tfrac12[\log_2(\lambda_{c,i}/\mu_c)]_+$ and $D_c=\tfrac{1}{n}\sum_i\min\{\lambda_{c,i},\mu_c\}$ for some component-specific water level $\mu_c$. The genie-aided conditional benchmark in which the active label is known at both terminals is
\begin{align}
\label{eq:Rcond}
R_{\rm cond}(D) \;\triangleq\; \inf_{\{D_c\ge 0\}_{c=1}^K:\,\sum_c \pi_c D_c\le D}\; \sum_{c=1}^{K}\pi_c R_c(D_c).
\end{align}

\begin{theorem}[Single global water level]
\label{thm:water}
The optimizer of \eqref{eq:Rcond} is characterized by a single global water level $\mu\ge 0$, and the conditional benchmark admits the explicit parametric form
\begin{align}
R_{\rm cond}(D) &= \frac{1}{n}\sum_{c=1}^{K}\pi_c\sum_{i=1}^{n} \frac{1}{2}\Bigl[\log_2\!\frac{\lambda_{c,i}}{\mu}\Bigr]_+, \label{eq:Rcond_explicit} \\ 
D &= \frac{1}{n}\sum_{c=1}^{K}\pi_c\sum_{i=1}^{n}\min\{\lambda_{c,i},\mu\}.
\label{eq:Dcond_explicit}
\end{align}
$R_{\rm cond}$ is non-increasing in $\mu$; $D$ is non-decreasing.
\end{theorem}

\begin{figure}[t]
\centering
\includegraphics[width=0.55\linewidth]{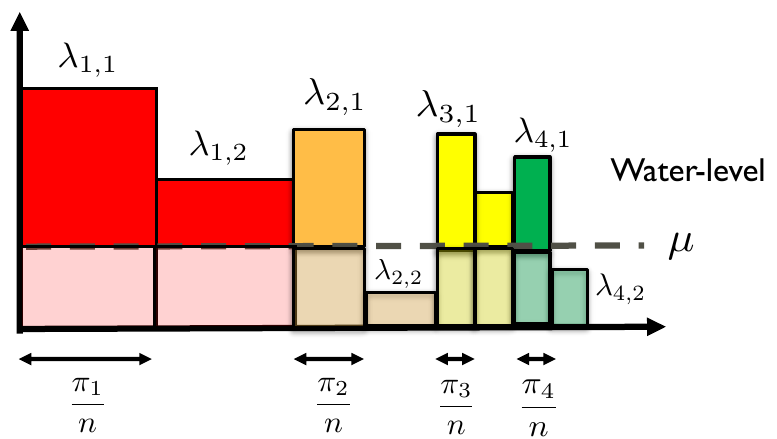}
\caption{Reverse waterfilling for a Gaussian-mixture source ($K=4$, $n=2$). Each column is one transform mode $(c,i)$ with height $\lambda_{c,i}$ and width $\pi_c/n$; the dashed line is the single
global water level $\mu$. Saturated caps above the line give the rate $r_{c,i}^\star = [\tfrac12\log_2(\lambda_{c,i}/\mu)]_+$; submerged portions give the distortion $d_{c,i}^\star=\min\{\lambda_{c,i},\mu\}$; eigenmodes $\lambda_{2,2}$ and $\lambda_{4,2}$ are pruned.}
\label{fig:bathtub}
\end{figure}

The proof (Appendix~\ref{app:proof_water}) is a Lagrangian KKT argument: because $\pi_c$ multiplies both the rate and the distortion contributions of $(c,i)$, it cancels in the stationarity condition, leaving a single common $\mu$ across all components and eigenmodes. The interpretation is that the multimodal allocation collapses onto exactly the single-Gaussian reverse-waterfilling rule, applied to the \emph{pooled} eigenmodes weighted by $\pi_c$.

\subsection{Converse--achievability sandwich}

\begin{theorem}[Genie-aided converse]
\label{thm:converse}
Let $R^\star(D)$ denote the operational RD function of the Gaussian-mixture source \eqref{eq:mixture}. Then
\begin{align}
\label{eq:converse}
R^\star(D) \ge R_{\rm cond}(D).
\end{align}
\end{theorem}

The converse follows because revealing the component $C$ to both terminals as free side information cannot increase the minimum required rate; see Appendix~\ref{app:proof_converse}.

\begin{theorem}[Label-aware achievability]
\label{thm:ach}
For the i.i.d. Gaussian-mixture source \eqref{eq:mixture}, \textsc{PrismQuant} achieves
\begin{align}
\label{eq:ach}
R_{\rm PQ}(D) \le R_{\rm cond}(D) + \frac{H(C)}{n}.
\end{align}
\end{theorem}

The proof (Appendix~\ref{app:proof_ach}) builds an explicit two-stage code that sends the label losslessly and then runs Shannon's lossy source-coding theorem on the conditionally Gaussian residual; the excess rate over $R_{\rm cond}(D)$ is exactly the label rate.

\begin{corollary}[Sandwich and asymptotic optimality]
\label{cor:sandwich}
For every $D\ge 0$,
\begin{align}
R_{\rm cond}(D) \;\le\; R^\star(D) \;\le\; R_{\rm cond}(D) + \frac{H(C)}{n} \;\le\; R_{\rm cond}(D) + \frac{\log_2 K}{n}.
\label{eq:sandwich}
\end{align}
If the mixture order is allowed to depend on the source dimension, i.e., $K=K_n$, and
\begin{align}
\label{eq:subexp}
\frac{\log_2 K_n}{n}\;\to\;0\qquad\text{as }n\to\infty,
\end{align}
then $R^\star(D)-R_{\rm cond}(D)\to 0$. Condition \eqref{eq:subexp} is strictly weaker than fixed $K$: it admits any sub-exponential growth $K_n=2^{o(n)}$, including polynomial growth $K_n=n^p$ and
$K_n=2^{n^{\alpha}}$ with $\alpha<1$. For $H(C)=0$ the sandwich is exact for all $n$ and $D$.
\end{corollary}

\paragraph{Why one water level suffices: the bathtub view.}
For a multimodal source, the textbook intuition is one component-specific water level per component plus another to balance the component budgets. Theorem~\ref{thm:water} shows that this is unnecessary: every extra bit spent on component-eigenmode pair $(c,i)$ is both \emph{paid} and \emph{rewarded} with probability $\pi_c$, so the prior cancels in the KKT condition and a single Lagrange multiplier $\mu$ governs the entire mixture. Figure~\ref{fig:bathtub} visualizes this as a bathtub: each $(c,i)$ is a rectangular column of height $\lambda_{c,i}$ and width $\pi_c/n$, and a single horizontal line at height $\mu$ decides simultaneously (i)~the active set across all $K$ components, (ii)~the per-eigenmode distortion $d_{c,i}^\star=\min\{\lambda_{c,i},\mu\}$ (equalized to $\mu$ on active eigenmodes by KKT), and (iii)~the per-eigenmode rate $r_{c,i}^\star =  [\tfrac12\log_2(\lambda_{c,i}/\mu)]_+$. The same $\mu$ arising in the conditional benchmark also determines \textsc{PrismQuant}'s on-line allocation, so the codec is parameterized by a single scalar dial regardless of $K$.

\paragraph{MAP label error in high dimension.}
Practical \textsc{PrismQuant} replaces the oracle label with a hard MAP decision $\hat C(x)=\argmax_c \pi_c \mathcal N(x;\mu_c,R_c)$. The standard Bhattacharyya union bound gives $P_e^{\rm MAP}\le \sum_{c<j}\sqrt{\pi_c\pi_j}\,\rho_{c,j}^{(n)}$ with
$\rho_{c,j}^{(n)}=\int\sqrt{f_c f_j}\,dx$ where $f_c$ denotes the conditional density for component $c \in [K]$. For Gaussian components, $\rho_{c,j}^{(n)}=\exp(-B_{c,j}^{(n)})$ where $B_{c,j}^{(n)}$ is the pairwise Bhattacharyya distance; closed-form expressions for $\rho_{c,j}^{(n)}$ across Gaussian, factorized Laplacian, uniform, and hybrid mixture families are provided in \citet{LeeEntropyMix2026}. Whenever the \emph{per-dimension Bhattacharyya distance} is bounded below by a positive constant (equivalently, components remain pairwise separable in the per-dimension limit), $P_e^{\rm MAP}$ decays exponentially in $n$ and the practical codec recovers the oracle-label benchmark of Theorem~\ref{thm:ach}. Consequently, for sufficiently high-dimensional sources, the MAP label error probability decays exponentially with dimension, causing practical \textsc{PrismQuant} to approach the genie-aided benchmark. More details are provided in Appendix \ref{app:map}.

\begin{figure}[t]

    \centering
    \begin{subfigure}[t]{0.3\textwidth}
        \includegraphics[width=\textwidth]{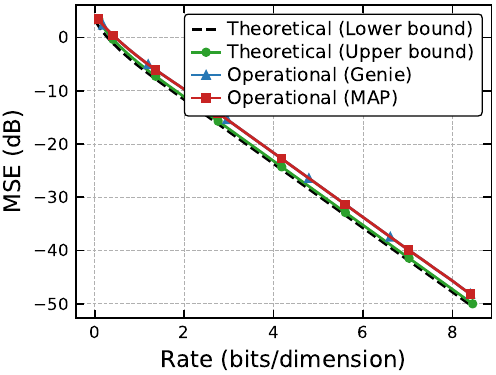}
        \caption{$K=8$}
    \end{subfigure}\hfill
    \begin{subfigure}[t]{0.3\textwidth}
        \includegraphics[width=\textwidth]{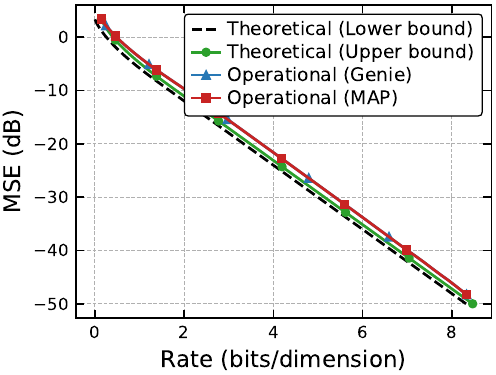}
        \caption{$K=32$}
    \end{subfigure}\hfill
    \begin{subfigure}[t]{0.3\textwidth}
        \includegraphics[width=\textwidth]{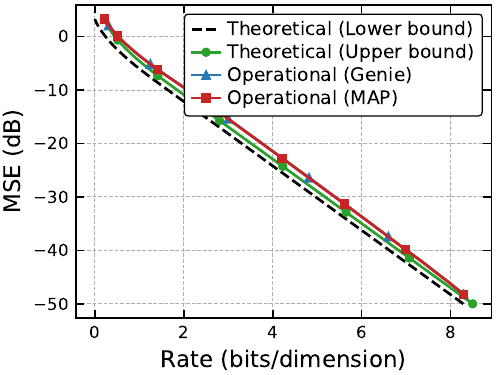}
        \caption{$K=128$}
    \end{subfigure}
    \caption{Effect of mixture order $K$ at fixed $n=32$. Operational (Genie) closely tracks the theoretical upper bound, validating Theorem~\ref{thm:water}; Operational (MAP) nearly coincides with Genie as $K \rightarrow \infty$.}
    \label{fig:synth_K}
\end{figure}

\section{Experiments}
\label{sec:csi}

\paragraph{Setup.}
We evaluate \textsc{PrismQuant} on synthetic mixture sources and on the DeepMIMO \citep{alkhateeb2019deepmimo} CSI dataset. For synthetic sources, weights $\{\pi_c\}_{c=1}^K$ are drawn uniformly from the simplex; component variances are drawn from $[0.1,10.0]$; means are zero; and $U_c$ is the QR factor of a random Gaussian matrix. For each configuration, more than 100,000 samples are drawn from the mixture and rate--distortion is swept by varying $\mu$ over 50 logarithmically-spaced points in $[10^{-5},10^{1}]$.  

For the DeepMIMO evaluation the dimensionality of CSI is parameterized with the number of antennas $N_t$ and the number of subcarriers $N_c$. Since the stacked CSI vector $h\in\mathbb{C}^{N_tN_c}$ is too high-dimensional for direct GMM fitting via EM, we apply a preprocessing step that first separates the real and imaginary parts of $h$ and concatenates them into a real-valued vector of length $2N_tN_c$, which is then partitioned into non-overlapping sub-vectors of length $n$, yielding $\lceil 2N_tN_c/n\rceil$ sub-vectors per CSI realization. Each $n$-dimensional sub-vector $x\in\mathbb{R}^{n}$ is then modeled by the Gaussian mixture. The dictionary $\theta=\{(\pi_c,\mu_c,R_c)\}_{c=1}^K$ is learned by EM
on the training split and frozen for inference; the label is amortized
over a $\tau$-vector coherence window, and since users in DeepMIMO
remain stationary we take $\tau \to \infty$, so that the label cost
vanishes and the effective rate reduces to $H(C)/(\tau n) \to 0$.

We compare four methods: the \textbf{Theoretical (lower bound)} $R_{\rm cond}(D)$ (Theorem~\ref{thm:converse}, computed via reverse waterfilling without label overhead), the \textbf{Theoretical (upper bound)} $R_{\rm PQ}(\mu)$ (Theorem~\ref{thm:ach}), the \textbf{Operational (Genie)} curve obtained by running \textsc{PrismQuant} in Section \ref{sec:prismquant} with oracle labels, and the \textbf{Operational (MAP)} curve obtained with MAP-inferred labels. Additional EM implementation details are deferred to Appendix~\ref{app:em}.

\paragraph{Effect of $K$ on synthetic mixtures.}
Figure~\ref{fig:synth_K} shows rate--distortion curves at fixed source dimension $n=32$ and mixture orders $K\in\{8,32,128\}$. Operational (Genie) closely tracks the theoretical upper bound, demonstrating that the global reverse-waterfilling structure of Theorem~\ref{thm:water} is directly achievable in practice through \textsc{PrismQuant}. As $K$ increases, the gap between the theoretical lower and upper bounds also increases due to the growing component-label cost $H(C)/n$, consistent with Theorem~\ref{thm:ach}. Nevertheless, even at $K=128$, the gap remains below $0.1$ bits/dimension, suggesting that the asymptotic gap rapidly vanishes in high dimension, consistent with Corollary~1. Moreover, Operational (MAP) nearly coincides with the Genie-aided scheme across all $K$, showing that MAP label estimation remains highly accurate even for large mixture orders. Additional experiments with varying source dimension $n$ are provided in Appendix~\ref{app:exp_extra}.

\begin{figure}[t]
    \centering
    \begin{subfigure}[t]{0.35\textwidth}
    \raisebox{3mm}{
        \hspace{0.08\textwidth}
        \includegraphics[width=\textwidth]{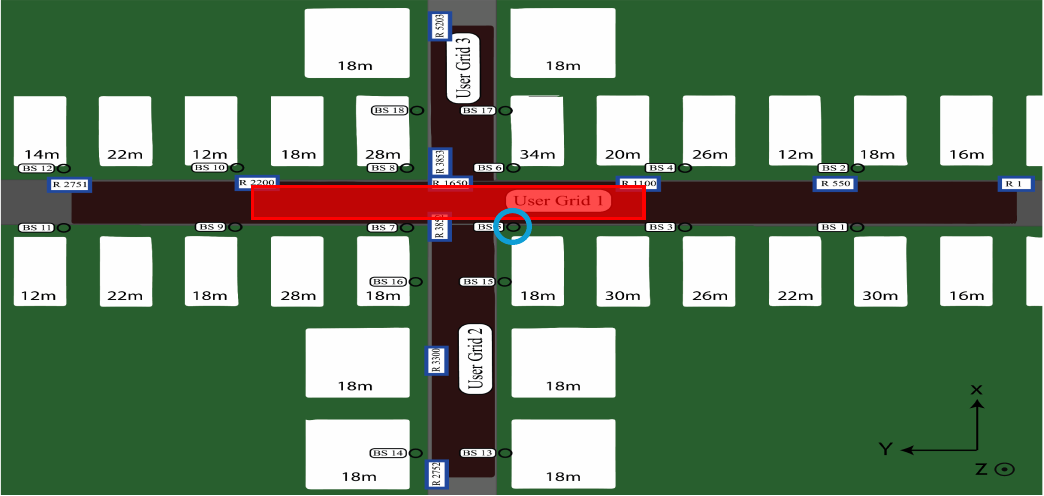}
    }
        \caption{Outdoor map}
    \end{subfigure}\hfill
    \begin{subfigure}[t]{0.4\textwidth}
        \includegraphics[width=\textwidth]{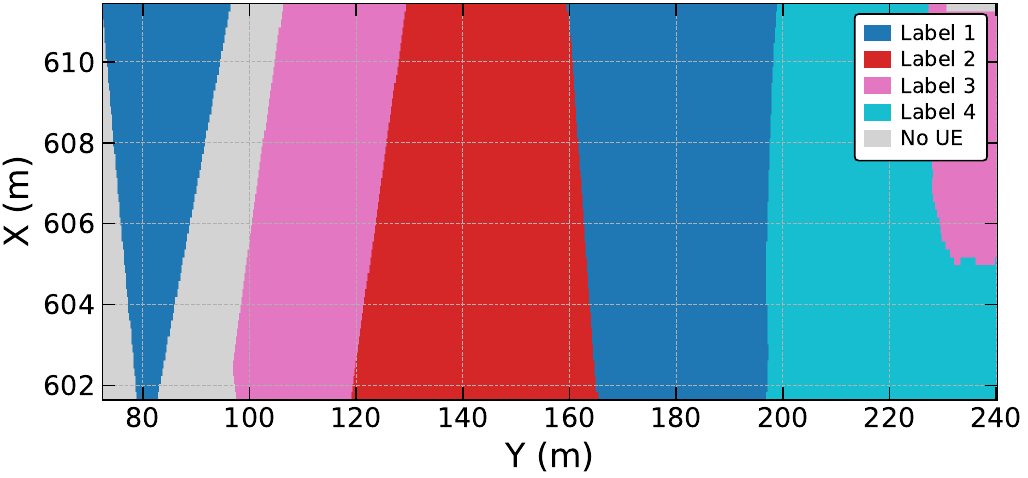}
        \caption{$K=4$}
    \end{subfigure}
    \vfill
    \begin{subfigure}[t]{0.4\textwidth}
        \includegraphics[width=\textwidth]{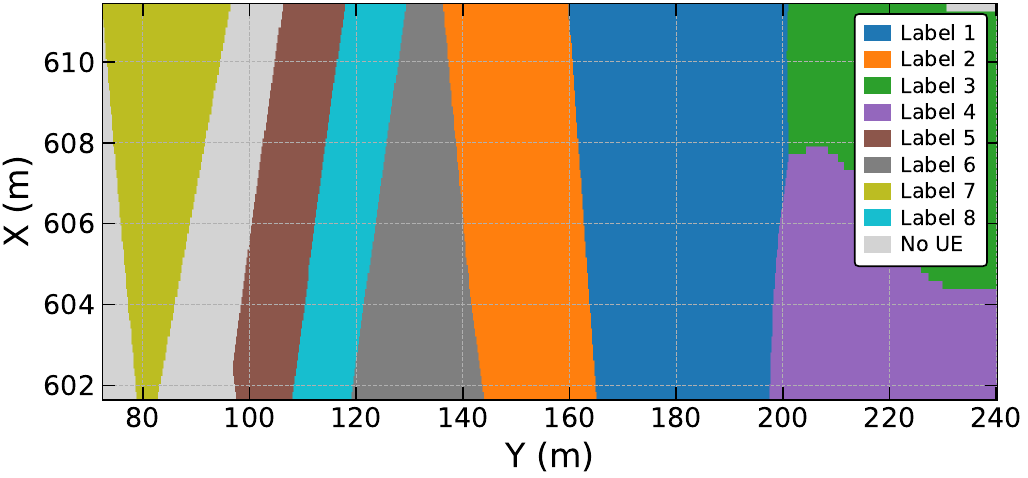}
        \caption{$K=8$}
    \end{subfigure}\hfill
    \begin{subfigure}[t]{0.4\textwidth}
        \includegraphics[width=\textwidth]{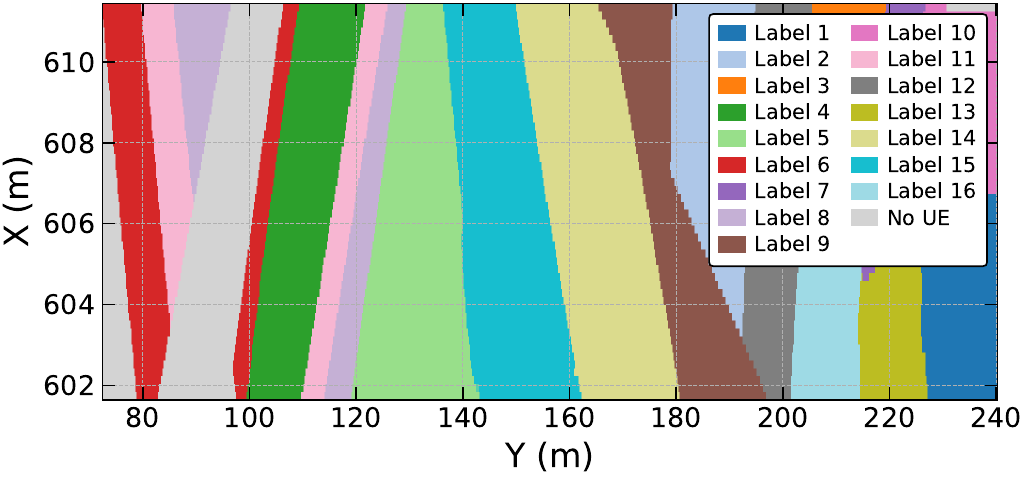}
        \caption{$K=16$}
    \end{subfigure}
   \caption{Spatial visualization of \textsc{PrismQuant}'s clustering on the DeepMIMO O1 dataset. (a) The outdoor deployment map used for CSI generation. (b)--(d) MAP component assignments---obtained from the EM-fitted Gaussian-mixture dictionary---plotted at each user's location, with varying $K$. Pixels sharing the same color correspond to CSI samples assigned to the same component.}    \label{fig:deepMIMO_map}
\end{figure}

\begin{figure}[t]
\centering
\begin{subfigure}{0.4\textwidth}
    \centering
    \includegraphics[width=\textwidth]{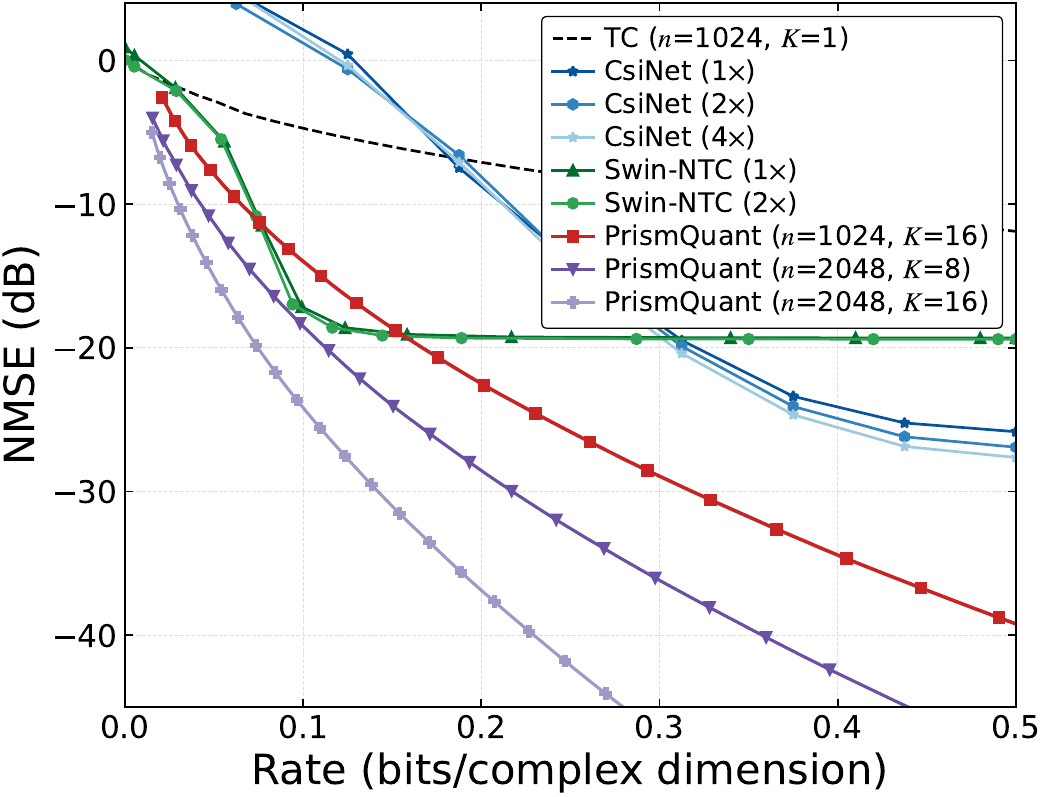}
    \caption{Comparison with baselines.}
    \label{fig:DeepMIMO_baselines}
\end{subfigure}
\hfill
\begin{subfigure}{0.4\textwidth}
    \centering
    \includegraphics[width=\textwidth]{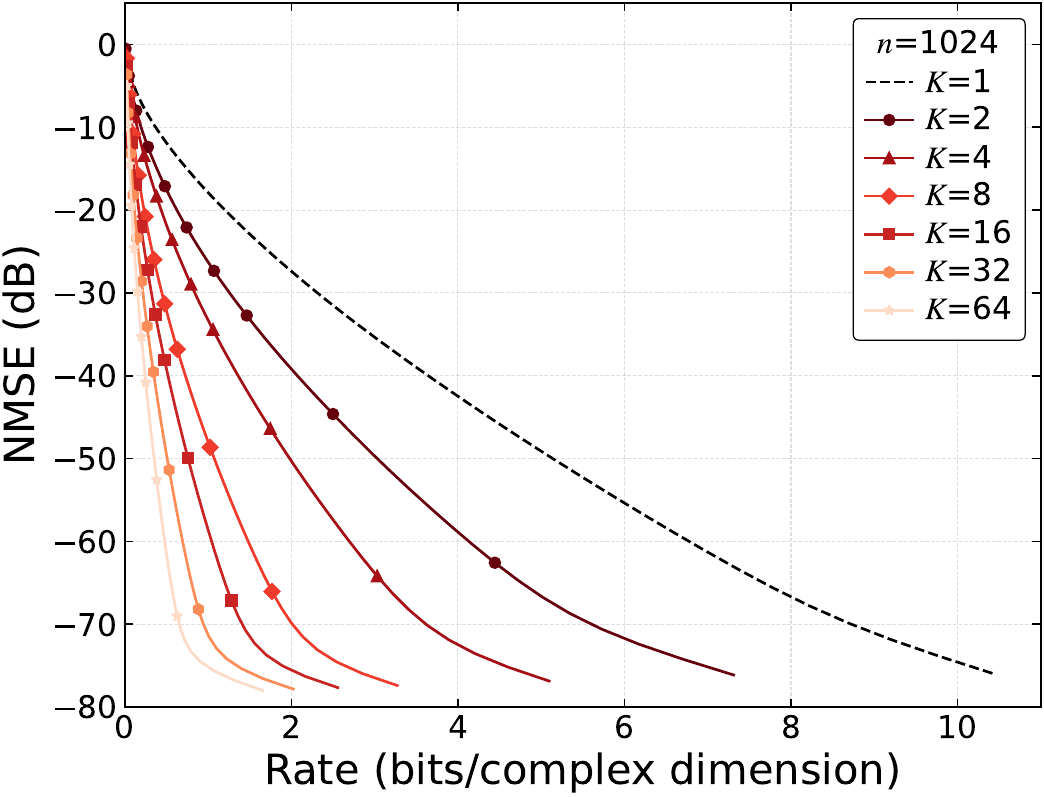}
    \caption{Effect of mixture order $K$.}
    \label{fig:DeepMIMO_K_sweep}
\end{subfigure}
\caption{RD performance on the DeepMIMO O1 dataset. (a)
\textsc{PrismQuant} compared with
single-covariance TC ($n{=}1024$, $K{=}1$),
Swin-NTC~\citep{Park25ICCNTC}, and CsiNet with uniform quantization
\citep{Wen18CsiNet}. (b) RD performance of \textsc{PrismQuant} at
$n{=}1024$ as the mixture order $K$ is swept.}
\label{fig:DeepMIMO}
\end{figure}

\paragraph{From covariance clustering to spatial partitioning.}
Figure~\ref{fig:deepMIMO_map} (b)--(d) provides a visual sanity check of the modeling assumption underlying \textsc{PrismQuant}: that CSI samples
sharing similar second-order statistics also share similar geometry.
Although the EM procedure operates purely in the covariance domain
and is given no positional information, the resulting MAP assignments
form spatially contiguous regions. Increasing the mixture order $K$ 
splits each region into finer sub-regions; from a compression
standpoint, this sharpens the
rate--distortion behavior at the cost of a larger dictionary.

\begin{table}[t]
\centering
\caption{Model size and FLOPs on DeepMIMO. FLOPs measure the computational cost of compressing and reconstructing one CSI realization. Since \textsc{PrismQuant}'s complexity and FLOPs depend on the operating rate, we report values averaged over the rate range $0\text{--}0.5$ bits per dimension.}
\label{tab:complexity}
\small
\begin{tabular}{lcc}
\toprule
Method & Params & FLOPs \\
\midrule
TC ($n=1024$, $K=1$) & $1.0{\times}10^5$ & $5.5{\times}10^6$ \\
\textsc{PrismQuant ($n=1024$, $K=16$)}& $1.1{\times}10^6$ & $5.5{\times}10^6$ \\
CsiNet (1x) \citep{Wen18CsiNet} & $1.1{\times}10^6$ & $1.0{\times}10^7$ \\
Swin-NTC (1x) \citep{Park25ICCNTC} & $1.6{\times}10^6$ & $4.1{\times}10^{8}$ \\
\bottomrule
\end{tabular}
\end{table}

\paragraph{CSI compression on DeepMIMO.}
We evaluate \textsc{PrismQuant} on the DeepMIMO `O1' outdoor scenario at $28$~GHz \citep{alkhateeb2019deepmimo}, generated via ray-tracing in Remcom Wireless InSite; the deployment map is shown in Figure~\ref{fig:deepMIMO_map}(a). We configure the system with $N_t{=}32$ antennas and $N_c{=}32$ OFDM subcarriers, and retain the $38{,}298$ users closest to the BS to avoid near-zero-power channels dominating the empirical distribution. The dataset is split into training and test sets with an $80/20$ ratio, and all CSI matrices are normalized to unit power following standard neural CSI-feedback practice~\citep{Liu22Markov}. A $K{=}16$ Gaussian mixture is learned by EM on the training split and then frozen for inference. We compare \textsc{PrismQuant} against single-covariance TC, Swin-NTC \citep{Park25ICCNTC}, and CsiNet \citep{Wen18CsiNet}. Figure~\ref{fig:DeepMIMO} and Table~\ref{tab:complexity} show that \textsc{PrismQuant} consistently outperforms single-covariance TC, with the gap widening at higher rates due to improved modeling of the conditional source distribution. Compared with learned baselines, \textsc{PrismQuant} continues to improve at moderate-to-high rates while Swin-NTC and CsiNet saturate. At $0.4$~bpp, \textsc{PrismQuant} achieves $15$~dB and $10$~dB lower NMSE than Swin-NTC (1x) and CsiNet (1x), respectively, while requiring $74.5\times$ and $1.8\times$ lower computational complexity at matched parameter budgets. Figure~\ref{fig:DeepMIMO}(b) further shows that increasing the mixture order $K$ consistently improves RD performance by capturing the multimodal source distribution more accurately.

\section{Conclusion}
\label{sec:conclusion}
We presented \textsc{PrismQuant}, a constructive transform-coding framework for Gaussian-mixture sources. By augmenting the source with the latent component label, we derived an explicit converse--achievability sandwich with gap $H(C)/n$ and showed that the entire multimodal RD problem is governed by a single global reverse-waterfilling level shared across all components and eigenmodes. We further developed a fully practical coding pipeline based on EM-driven Gaussian-mixture learning, MAP component inference, component-matched KLTs, and ECSQ. Experiments on synthetic Gaussian mixtures demonstrated that \textsc{PrismQuant} closely approaches the theoretical RD bound, while experiments on the DeepMIMO CSI dataset showed competitive or superior performance compared with transformer-based learned codecs at two orders of magnitude smaller model size and complexity. Extending the experimental validation to broader multimodal source classes remains an important direction for future work.

% \textsc{PrismQuant} extends the explicit constructive RD theory of
% Gaussian transform coding to general Gaussian-mixture sources via an
% exact converse--achievability sandwich of gap $H(C)/n$, governed by a
% single global reverse-waterfilling level performing cross-component
% pruning under one dial. Via Gaussian-mixture universality the same
% machinery applies to arbitrary continuous multimodal sources through
% EM fitting; CSI feedback is the representative application.
% \textbf{Limitations.} The finite-$(K,n)$ overhead is upper-bounded by
% $\log_2 K/n$ (0.094~bit/dim at $K{=}64,n{=}64$); ECSQ adds a
% $\sim\!0.255$~bit/active-mode shaping loss.

% --- References (1 page) ---
\newpage
% References in \small (9 pt), per the NeurIPS 2026 style file.
{\small
\bibliographystyle{unsrtnat}
\bibliography{refs}

\begin{thebibliography}{39}
\providecommand{\natexlab}[1]{#1}
\providecommand{\url}[1]{\texttt{#1}}
\expandafter\ifx\csname urlstyle\endcsname\relax
  \providecommand{\doi}[1]{doi: #1}\else
  \providecommand{\doi}{doi: \begingroup \urlstyle{rm}\Url}\fi

\bibitem[Shannon(1959)]{Shannon59}
Claude~E. Shannon.
\newblock Coding theorems for a discrete source with a fidelity criterion.
\newblock \emph{IRE National Convention Record}, 7\penalty0 (4):\penalty0
  142--163, 1959.

\bibitem[Berger(1971)]{Berger71}
Toby Berger.
\newblock \emph{Rate distortion theory: A mathematical basis for data
  compression}.
\newblock Prentice-Hall, Englewood Cliffs, NJ, 1971.

\bibitem[Cover and Thomas(2006)]{CoverThomas06}
Thomas~M. Cover and Joy~A. Thomas.
\newblock \emph{Elements of information theory}.
\newblock Wiley--Interscience, New York, NY, 2nd edition, 2006.

\bibitem[Goyal(2001)]{Goyal01}
Vivek~K. Goyal.
\newblock Theoretical foundations of transform coding.
\newblock \emph{IEEE Signal Procesing Magazine}, 18\penalty0 (5):\penalty0
  9--21, 2001.

\bibitem[Bishop(2006)]{Bishop06}
Christopher~M. Bishop.
\newblock \emph{Pattern recognition and machine learning}.
\newblock Springer, New York, NY, 2006.

\bibitem[Marzetta(2010)]{Marzetta10MM}
T.~L. Marzetta.
\newblock Noncooperative cellular wireless with unlimited numbers of base
  station antennas.
\newblock \emph{IEEE Transactions on Wireless Communications}, 9\penalty0
  (11):\penalty0 3590--3600, 2010.

\bibitem[Love et~al.(2008)Love, {Heath, Jr.}, Lau, Gesbert, Rao, and
  Andrews]{Love08LimitedFB}
D.~J. Love, R.~W. {Heath, Jr.}, V.~K.~N. Lau, D.~Gesbert, B.~D. Rao, and
  M.~Andrews.
\newblock An overview of limited feedback in wireless communication systems.
\newblock \emph{IEEE Journal on Selected Areas in Communications}, 26\penalty0
  (8):\penalty0 1341--1365, 2008.

\bibitem[{Heath, Jr.} and Lozano(2016)]{Heath16MIMO}
R.~W. {Heath, Jr.} and A.~Lozano.
\newblock \emph{Foundations of {MIMO} Communication}.
\newblock Cambridge Univ. Press, 2016.

\bibitem[Liu et~al.(2012)Liu, Oestges, Poutanen, Quitin, Haneda, Vainikainen,
  Tufvesson, Molisch, and Doncker]{Liu12COST2100}
Lingfeng Liu, Claude Oestges, Juho Poutanen, Francois Quitin, Katsuyuki Haneda,
  Pekka Vainikainen, Fredrik Tufvesson, Andreas~F. Molisch, and Pierre~De
  Doncker.
\newblock The {COST} 2100 {MIMO} channel model.
\newblock \emph{IEEE Wireless Communications}, 19\penalty0 (6):\penalty0
  92--99, 2012.

\bibitem[Wen et~al.(2018)Wen, Shih, and Jin]{Wen18CsiNet}
C.-K. Wen, W.-T. Shih, and S.~Jin.
\newblock Deep learning for massive {MIMO} {CSI} feedback.
\newblock \emph{IEEE Wireless Commun. Lett.}, 7\penalty0 (5):\penalty0
  748--751, 2018.

\bibitem[Park et~al.(2025)Park, Do, and Lee]{Park25ICCNTC}
B.~Park, H.~Do, and N.~Lee.
\newblock Transformer-based nonlinear transform coding for multi-rate {CSI}
  compression in {MIMO}-{OFDM} systems.
\newblock In \emph{Proceedings of IEEE International Conference on
  Communications (ICC)}, 2025.

\bibitem[Lee(2026)]{LeeEntropyMix2026}
Namyoon Lee.
\newblock On the entropy of general mixture distributions, 2026.
\newblock arXiv:2602.15303.

\bibitem[Effros and Chou(1995)]{EffrosChou95}
M.~Effros and P.~A. Chou.
\newblock Weighted universal transform coding: Universal image compression with
  the {K}arhunen--{L}o\`eve transform.
\newblock In \emph{Proceedings of IEEE International Conference on Image
  Processing (ICIP)}, 1995.

\bibitem[Alkhateeb(2019)]{alkhateeb2019deepmimo}
Ahmed Alkhateeb.
\newblock Deep{MIMO}: A generic deep learning dataset for millimeter wave and
  massive {MIMO} applications, 2019.
\newblock arXiv:1902.06435.

\bibitem[Gray(1972)]{Gray72CondRD}
Robert~M. Gray.
\newblock Conditional rate--distortion theory.
\newblock Technical report, Stanford Electronics Laboratories, Stanford
  University, 1972.

\bibitem[Wyner and Ziv(1976)]{WynerZiv76}
A.~D. Wyner and J.~Ziv.
\newblock The rate--distortion function for source coding with side information
  at the decoder.
\newblock \emph{IEEE Transactions on Information Theory}, 22\penalty0
  (1):\penalty0 1--10, 1976.

\bibitem[Heegard and Berger(1985)]{HeegardBerger85}
C.~Heegard and T.~Berger.
\newblock Rate distortion when side information may be absent.
\newblock \emph{IEEE Transactions on Information Theory}, 31\penalty0
  (6):\penalty0 727--734, 1985.

\bibitem[Riskin and Gray(1991)]{RiskinGray91}
E.~A. Riskin and R.~M. Gray.
\newblock A greedy tree growing algorithm for the design of variable rate
  vector quantizers.
\newblock \emph{IEEE Transactions on Signal Processing}, 39\penalty0
  (11):\penalty0 2500--2507, 1991.

\bibitem[Effros et~al.(1994)Effros, Chou, and Gray]{EffrosChouGray94}
M.~Effros, P.~A. Chou, and R.~M. Gray.
\newblock Variable dimension weighted universal vector quantization and
  noiseless coding.
\newblock In \emph{Proceedings of IEEE Data Compression Conference (DCC)},
  1994.

\bibitem[Linde et~al.(1980)Linde, Buzo, and Gray]{LindeBuzoGray80}
Y.~Linde, A.~Buzo, and R.~M. Gray.
\newblock An algorithm for vector quantizer design.
\newblock \emph{IEEE Transactions on Communications}, 28\penalty0 (1):\penalty0
  84--95, 1980.

\bibitem[Hedelin and Skoglund(2000)]{HedelinSkoglund00}
P.~Hedelin and J.~Skoglund.
\newblock Vector quantization based on {G}aussian mixture models.
\newblock \emph{IEEE Transactions on Speech and Audio Processing}, 8\penalty0
  (4):\penalty0 385--401, 2000.

\bibitem[Han et~al.(2010)Han, Saxena, and Rose]{HanSaxenaRose10ASTC}
J.~Han, A.~Saxena, and K.~Rose.
\newblock Towards jointly optimal spatial prediction and adaptive transform in
  video/image coding.
\newblock In \emph{Proc. IEEE Int. Conf. Acoust., Speech, Signal Process.
  (ICASSP)}, pages 726--729, Mar. 2010.

\bibitem[Han et~al.(2012)Han, Saxena, Melkote, and
  Rose]{HanSaxenaMelkoteRose12AMT}
J.~Han, A.~Saxena, V.~Melkote, and K.~Rose.
\newblock Jointly optimized spatial prediction and block transform for video
  and image coding.
\newblock \emph{IEEE Trans. Image Process.}, 21\penalty0 (4):\penalty0
  1874--1884, Apr. 2012.

\bibitem[Saxena and Fernandes(2013)]{SaxenaFernandes13AMT}
A.~Saxena and F.~C. Fernandes.
\newblock {DCT}/{DST}-based transform coding for intra prediction in
  image/video coding.
\newblock \emph{IEEE Trans. Image Process.}, 22\penalty0 (10):\penalty0
  3974--3981, Oct. 2013.

\bibitem[Zhao et~al.(2016)Zhao, Chen, Karczewicz, Said, and Seregin]{ZhaoEMT16}
X.~Zhao, J.~Chen, M.~Karczewicz, A.~Said, and V.~Seregin.
\newblock Enhanced multiple transform for video coding.
\newblock In \emph{Proc. Data Compression Conf. (DCC)}, pages 73--82, Mar.
  2016.

\bibitem[Bross et~al.(2021)Bross, Wang, Ye, Liu, Chen, Sullivan, and
  Ohm]{BrossVVC21}
B.~Bross, Y.-K. Wang, Y.~Ye, S.~Liu, J.~Chen, G.~J. Sullivan, and J.-R. Ohm.
\newblock Overview of the {V}ersatile {V}ideo {C}oding ({VVC}) standard and its
  applications.
\newblock \emph{IEEE Trans. Circuits Syst. Video Technol.}, 31\penalty0
  (10):\penalty0 3736--3764, Oct. 2021.

\bibitem[Ball{\'e} et~al.(2018)Ball{\'e}, Minnen, Singh, Hwang, and
  Johnston]{Balle18Hyperprior}
Johannes Ball{\'e}, David Minnen, Saurabh Singh, Sung~Jin Hwang, and Nick
  Johnston.
\newblock Variational image compression with a scale hyperprior.
\newblock In \emph{Proceedings of International Conference on Learning
  Representations (ICLR)}, 2018.

\bibitem[Cheng et~al.(2020)Cheng, Sun, Takeuchi, and Katto]{cheng2020learned}
Zhengxue Cheng, Heming Sun, Masaru Takeuchi, and Jiro Katto.
\newblock Learned image compression with discretized gaussian mixture
  likelihoods and attention modules.
\newblock In \emph{Proceedings of the IEEE/CVF Conference on Computer Vision
  and Pattern Recognition (CVPR)}, 2020.

\bibitem[Lei et~al.(2022)Lei, Hassani, and Saeedi~Bidokhti]{10124059}
Eric Lei, Hamed Hassani, and Shirin Saeedi~Bidokhti.
\newblock Neural estimation of the rate-distortion function with applications
  to operational source coding.
\newblock \emph{IEEE Journal on Selected Areas in Information Theory},
  3\penalty0 (4):\penalty0 674--686, 2022.

\bibitem[Zandieh et~al.(2025)Zandieh, Daliri, Hadian, and
  Mirrokni]{Zandieh2025TurboQuant}
Amir Zandieh, Majid Daliri, Majid Hadian, and Vahab Mirrokni.
\newblock Turbo{Q}uant: Online vector quantization with near-optimal distortion
  rate, 2025.
\newblock arXiv:2504.19874.

\bibitem[Dempster et~al.(1977)Dempster, Laird, and Rubin]{Dempster77}
A.~P. Dempster, N.~M. Laird, and D.~B. Rubin.
\newblock Maximum likelihood from incomplete data via the {EM} algorithm.
\newblock \emph{Journal on Royal Statistical Society: Series B
  (Methodological)}, 39\penalty0 (1):\penalty0 1--22, 1977.

\bibitem[Wen et~al.(2015)Wen, Jin, Wong, Chen, and Ting]{wen2015channel}
Chao-Kai Wen, Shi Jin, Kai-Kit Wong, Jung-Chieh Chen, and Pangan Ting.
\newblock Channel estimation for massive {MIMO} using {G}aussian-mixture
  bayesian learning.
\newblock \emph{IEEE Transactions on Wireless Communications}, 14\penalty0
  (3):\penalty0 1356--1368, 2015.

\bibitem[Song et~al.(2025)Song, Yang, Barzegar~Khalilsarai, and
  Caire]{song2025downlink}
Yi~Song, Tianyu Yang, Mahdi Barzegar~Khalilsarai, and Giuseppe Caire.
\newblock Downlink {CSIT} under compressed feedback: Joint versus separate
  source-channel coding.
\newblock \emph{IEEE Transactions on Wireless Communications}, 24\penalty0
  (10):\penalty0 8429--8444, 2025.

\bibitem[Nguyen et~al.(2005)Nguyen, Worring, and Smeulders]{nguyen2005study}
Hieu~T. Nguyen, Marcel Worring, and Arnold W.~M. Smeulders.
\newblock A study of {G}aussian mixture models of color and texture features
  for image classification and segmentation.
\newblock \emph{Pattern Recognition}, 38\penalty0 (11):\penalty0 2005--2018,
  2005.

\bibitem[Zoran and Weiss(2011)]{zoran2011learning}
Daniel Zoran and Yair Weiss.
\newblock From learning models of natural image patches to whole image
  restoration.
\newblock In \emph{Proceedings of IEEE International Conference on Computer
  Vision (ICCV)}, 2011.

\bibitem[Falck et~al.(2021)Falck, Zhang, Willetts, Nicholson, Yau, and
  Holmes]{falck2021multi}
Fabian Falck, Haoting Zhang, Matthew Willetts, George Nicholson, Christopher
  Yau, and Chris~C Holmes.
\newblock Multi-facet clustering variational autoencoders.
\newblock \emph{Advances in Neural Information Processing Systems},
  34:\penalty0 8676--8690, 2021.

\bibitem[Povey et~al.(2010)Povey, Burget, Agarwal, Akyazi, Kai, Ghoshal,
  Glembek, Goel, Karafiat, Rastrow, et~al.]{povey2010subspace}
Daniel Povey, Lukas Burget, Mohit Agarwal, Pinar Akyazi, Fan Kai, Arnab
  Ghoshal, Ondrej Glembek, Nagendra Goel, Martin Karafiat, Ariya Rastrow,
  et~al.
\newblock Subspace gaussian mixture models for speech recognition.
\newblock In \emph{Proceedings of IEEE International Conference on Acoustics,
  Speech and Signal Processing (ICASSP)}, 2010.

\bibitem[Gersho and Gray(1992)]{GershoGray92}
Allen Gersho and Robert~M. Gray.
\newblock \emph{Vector quantization and signal compression}.
\newblock Springer, Boston, MA, 1992.

\bibitem[Liu et~al.(2022)Liu, del Rosario, and Ding]{Liu22Markov}
Zhenyu Liu, Mason del Rosario, and Zhi Ding.
\newblock A {M}arkovian model-driven deep learning framework for massive {MIMO}
  {CSI} feedback.
\newblock \emph{IEEE Transactions on Wireless Communications}, 21\penalty0
  (2):\penalty0 1214--1228, 2022.

\end{thebibliography}
}

\newpage
\appendix

% =====================================================================
% Technical appendix for PrismQuant (CSI variant) NeurIPS 2026 submission.
% =====================================================================

\section{Proof of Theorem~\ref{thm:water} (single global water level)}
\label{app:proof_water}

The conditional benchmark \eqref{eq:Rcond} is a separable convex
program over $\{D_c\}$ subject to a single linear constraint. Form the
Lagrangian
\begin{align}
\mathcal L(\{D_c\},\nu,\{\eta_c\})
=\sum_{c=1}^{K}\pi_c R_c(D_c)
+ \nu\Bigl(\sum_{c=1}^{K}\pi_c D_c - D\Bigr)
- \sum_{c=1}^{K}\eta_c D_c,
\end{align}
with $\nu\ge 0$ and $\eta_c\ge 0$. KKT conditions: primal feasibility
$D_c\ge 0$, $\sum_c \pi_c D_c\le D$; dual feasibility; complementarity
$\nu(\sum_c\pi_c D_c-D)=0$, $\eta_c D_c=0$; stationarity
\begin{align}
\pi_c\frac{dR_c}{dD_c} + \nu\pi_c - \eta_c = 0,\qquad c\in[K].
\end{align}
For the Gaussian RD function on a fixed component $c$, on any interval
where the active set $A_c(\mu_c)=\{i:\lambda_{c,i}>\mu_c\}$ is constant,
\begin{align}
R_c(\mu_c)=\frac{1}{n}\sum_{i\in A_c(\mu_c)}\frac12\log_2\!\frac{\lambda_{c,i}}{\mu_c},
\quad
D_c(\mu_c)=\frac{1}{n}\Bigl(\sum_{i\not\in A_c(\mu_c)}\lambda_{c,i}+|A_c(\mu_c)|\mu_c\Bigr),
\end{align}
so
\begin{align}
\frac{dR_c}{dD_c}
=\frac{dR_c/d\mu_c}{dD_c/d\mu_c}
=-\frac{1}{\mu_c\ln 2}.
\end{align}
For any active component ($D_c>0$, hence $\eta_c=0$), stationarity
collapses to $\pi_c\bigl(-\tfrac{1}{\mu_c\ln 2}+\nu\bigr)=0$, i.e.
\begin{align}
\mu_c=\frac{1}{\nu\ln 2}\;\triangleq\;\mu,
\end{align}
which is independent of $c$. Hence a \emph{single} water level $\mu$
applies across all components and all eigenmodes. Substituting back,
\begin{align}
R_{\rm cond}(D)=\frac{1}{n}\sum_{c}\pi_c\sum_{i}\frac12\Bigl[\log_2\!\frac{\lambda_{c,i}}{\mu}\Bigr]_+,\quad
D=\frac{1}{n}\sum_c\pi_c\sum_i\min\{\lambda_{c,i},\mu\}.
\end{align}
Monotonicity is immediate: $\min\{\lambda_{c,i},\mu\}$ is nondecreasing
in $\mu$, so $D$ is nondecreasing; for $\mu_1<\mu_2$ the active set
$A_c(\mu_2)\subseteq A_c(\mu_1)$ and each summand decreases in $\mu$,
so $R_{\rm cond}$ is nonincreasing.

The interpretation is that the multimodal allocation collapses onto
the same single-Gaussian reverse-waterfilling rule applied to the
\emph{pooled} eigenmodes weighted by $\pi_c$: a mode $(c,i)$ is active
iff $\lambda_{c,i}>\mu$, regardless of $\pi_c$, while the prior enters
only through the expected-rate sum.

\section{Proof of Theorem~\ref{thm:converse} (genie-aided converse)}
\label{app:proof_converse}

Consider any admissible test channel $p(\hat x\mid x)$ with
$\E\|x-\hat x\|_2^2\le nD$. Since the latent state is generated first
and the reproduction depends on $C$ only through $x$, $C-x-\hat x$ is
a Markov chain. Then
\begin{align}
I(x;\hat x)
&= I(x,C;\hat x)-I(C;\hat x\mid x)
= I(C;\hat x)+I(x;\hat x\mid C)\\
&\ge I(x;\hat x\mid C)
=\sum_{c=1}^K \pi_c I(x;\hat x\mid C{=}c).
\end{align}
For each branch let $D_c\triangleq \E[\|x-\hat x\|^2\mid C{=}c]/n$. By
the definition of the per-branch Gaussian RD function,
$\tfrac{1}{n}I(x;\hat x\mid C{=}c)\ge R_c(D_c)$, so
\begin{align}
\frac{1}{n}I(x;\hat x)\;\ge\; \sum_{c=1}^K \pi_c R_c(D_c).
\end{align}
Since $\sum_c\pi_c D_c\le D$, the right-hand side is at least
$R_{\rm cond}(D)$. Taking the infimum over admissible test channels
gives $R^\star(D)\ge R_{\rm cond}(D)$.

\section{Proof of Theorem~\ref{thm:ach} (label-aware achievability)}
\label{app:proof_ach}

Fix $D\ge 0$ and $\epsilon>0$. By the definition of $R_{\rm cond}(D)$,
choose $\{D_c\}$ with $\sum_c\pi_c D_c\le D$ and
$\sum_c \pi_c R_c(D_c)\le R_{\rm cond}(D)+\epsilon/2$. For each branch
$c$, choose a conditional test channel $q_c(\hat x\mid x)$ with
$\E[\|x-\hat x\|^2\mid C{=}c]/n\le D_c$ and
$\tfrac{1}{n}I(x;\hat x\mid C{=}c)\le R_c(D_c)+\epsilon/2$. Define the
augmented source $Y=(C,x)$ with reproduction $\hat Y=(\hat C,\hat x)$
via
\begin{align}
q(\hat c,\hat x\mid c,x)=\mathbf 1\{\hat c=c\}\,q_c(\hat x\mid x).
\end{align}
The label is reproduced losslessly and the source by a branch-matched
channel; the total distortion is
$\E\|x-\hat x\|^2/n\le \sum_c \pi_c D_c\le D$. The mutual information
satisfies
\begin{align}
\frac{1}{n}I(Y;\hat Y)
&=\frac{H(C)}{n}+\frac{1}{n}I(x;\hat x\mid C)\\
&\le \frac{H(C)}{n}+\sum_c \pi_c R_c(D_c)+\frac{\epsilon}{2}
\le \frac{H(C)}{n}+R_{\rm cond}(D)+\epsilon.
\end{align}
Apply Shannon's lossy source-coding theorem to the augmented source
$Y=(C,x)$: for every sufficiently large blocklength there exists a
code with rate at most the right-hand side and expected distortion at
most $D+\epsilon$. Re-normalizing per source dimension gives
$R_{\rm PQ}(D)\le R_{\rm cond}(D)+H(C)/n+\epsilon$; letting
$\epsilon\downarrow 0$ proves the claim.  

\section{EM update for the Gaussian-mixture dictionary}
\label{app:em}

The dictionary $\{(\pi_c,\mu_c,R_c)\}_{c=1}^{K}$ is learned offline by
EM on a training set $\{x^{(j)}\}_{j=1}^{N_{\rm tr}}$ of source
vectors (real for Section~\ref{sec:model}; complex for the CSI
application of Section~\ref{sec:csi}, with $\mathcal N$ replaced by
$\mathcal{CN}$ in what follows). The E-step computes responsibilities
\begin{align}
r_{j,c}=\Pr(C=c\mid x^{(j)})=
\frac{\pi_c\,\mathcal N(x^{(j)};\mu_c,R_c)}{\sum_{d=1}^{K}\pi_d\,\mathcal N(x^{(j)};\mu_d,R_d)},
\end{align}
and the M-step updates the prior, mean, and covariance by weighted
moments:
\begin{align}
\pi_c^{\rm new}=\frac{1}{N_{\rm tr}}\sum_j r_{j,c},\quad
\mu_c^{\rm new}=\frac{\sum_j r_{j,c}\,x^{(j)}}{\sum_j r_{j,c}},\quad
R_c^{\rm new}=\frac{\sum_j r_{j,c}\,(x^{(j)}-\mu_c^{\rm new})(x^{(j)}-\mu_c^{\rm new})^{\rm H}}{\sum_j r_{j,c}}.
\end{align}
The KLT $U_c$ and eigenvalues $\Lambda_c$ follow from the spectral
decomposition $R_c^{\rm new}=U_c\Lambda_c U_c^{\rm H}$. Per-mode
Lloyd--Max scalar quantizers can be precomputed once $\Lambda_c$ is
known. For zero-mean sources (e.g., proper-complex CSI), the M-step
mean update collapses to $\mu_c^{\rm new}=0$.

\paragraph{Numerical stability of complex EM.}
For high-dimensional CSI, the empirical second-moment $R_c^{\rm new}$
can be ill-conditioned at small $K$ effective sample sizes
($\sum_j r_{j,c}\ll n$). We add a Tikhonov regularizer
$R_c^{\rm new}\!\leftarrow\!R_c^{\rm new}+\lambda I$ with
$\lambda{=}10^{-4}\,\tr(R_c^{\rm new})/n$, guaranteeing strict
positive definiteness; degenerate components (priors below
$10^{-4}$) are pruned and replaced by a split of the largest
component. Across $5$ random restarts on COST2100 the largest
across-restart NMSE deviation is $<0.1$~dB at every measured rate.

\section{MAP label error: closed-form bound for Gaussian components}
\label{app:map}

For Gaussian components, the pairwise Bhattacharyya overlap satisfies
$\rho_{c,j}^{(n)}=\exp(-B_{c,j}^{(n)})$ with
\begin{align}
B_{c,j}^{(n)}
=\frac{1}{8}(\mu_c-\mu_j)^{\rm H}\!\Bigl(\frac{R_c+R_j}{2}\Bigr)^{\!-1}\!(\mu_c-\mu_j)
+\frac{1}{2}\log\frac{\det\!\bigl(\tfrac{R_c+R_j}{2}\bigr)}
{\sqrt{\det R_c\,\det R_j}}.
\end{align}
The same Bhattacharyya overlap drives the closed-form bounds and
approximations of mixture entropy in \citet{LeeEntropyMix2026}, where
analogous expressions are derived for Gaussian, factorized Laplacian,
uniform, and hybrid mixture families; the bound below is the
Gaussian specialization used in the present paper.
When $\mu_c\ne\mu_j$, both the Mahalanobis term and the determinant
term contribute; for proper-complex CSI with $\mu_c=0$ only the
determinant term remains. The standard Bhattacharyya union bound
\begin{align}
P_e^{\rm MAP}\le \sum_{c<j}\sqrt{\pi_c\pi_j}\,e^{-B_{c,j}^{(n)}}
\le \frac{K-1}{2}\,e^{-\beta n}
\end{align}
holds whenever $B_{c,j}^{(n)}/n\ge\beta>0$ for all $c\ne j$, by
Cauchy--Schwarz on the prior square roots. Hence MAP errors decay
exponentially in $n$.

\section{Comparison with Classical and Standardized Transform Coding Methods}
\label{app:tc_wutc_pq}

This appendix places \textsc{PrismQuant} within a four-tier hierarchy of transform-based source codes: classical single-covariance transform coding (TC), the learned-library Weighted Universal Transform Coder (WUTC) of \citet{EffrosChou95}, the standardized Adaptive Multiple Transform (AMT/MTS) of HEVC and VVC \citep{HanSaxenaRose10ASTC,HanSaxenaMelkoteRose12AMT,SaxenaFernandes13AMT,ZhaoEMT16,BrossVVC21}, and \textsc{PrismQuant}. All four schemes share the same encoder skeleton---a library of orthogonal transforms with (possibly trivial) branch signaling---but differ in how the library is designed, how the branch is selected, how bits are allocated, and what information-theoretic guarantees are available. We first lay out the side-by-side architectural comparison, then prove \emph{information-theoretically} why a single global water level strictly dominates per-class water levels at every operating point, then specialize the discussion to AMT, and finally confirm the analysis on a synthetic mixture.

\subsection{Four architectures, side by side}

\begin{table}[h]
\centering
\caption{Single-covariance TC, WUTC \citep{EffrosChou95}, AMT/MTS \citep{SaxenaFernandes13AMT,BrossVVC21}, and \textsc{PrismQuant}: same outer pipeline; the library, the branch-selection rule, and the bit-allocation rule differ.}
\label{tab:tc_wutc_pq}
\small
\renewcommand{\arraystretch}{1.25}
\setlength{\tabcolsep}{4pt}
\newcolumntype{Y}{>{\raggedright\arraybackslash}X}
\begin{tabularx}{\linewidth}{@{}l Y Y Y Y@{}}
\toprule
 & \textbf{Single-cov TC} & \textbf{WUTC} & \textbf{AMT / MTS} & \textbf{\textsc{PrismQuant}} \\
\midrule
Transform library
 & one global KLT $U$
 & data-driven $\{U_c\}_{c=1}^{K}$ via LBG
 & fixed trigonometric library (DCT-II, DST-VII, DCT-VIII)
 & GMM-induced $\{U_c\}_{c=1}^{K}$ as eigendecomp.\ of $\Sigma_c$ \\
Source model
 & single Gaussian $\mathcal N(0,\Sigma)$
 & none (operational)
 & none (universal)
 & explicit GMM $\sum_c\pi_c\mathcal N(\mu_c,\Sigma_c)$ \\
Label transmitted?
 & no
 & yes (lossless, entropy-coded)
 & yes (1--3 bits/block, CABAC)
 & yes (lossless at rate $H(\widehat C)/n$) \\
Branch selection rule
 & N/A
 & encoder-side RD search over $K$
 & encoder-side RD search over $\sim 5$
 & \textbf{Bayes-optimal MAP} $\widehat C(x)$ \\
Scalar quantizer
 & ECSQ on $U^{\!\top}\!X$
 & ECSQ on $U_{\hat c}^{\!\top}\!X$
 & per-block QP from standard table
 & ECSQ on $U_{\hat c}^{\!\top}\!X$ \\
\textbf{Bit allocation}
 & global $\mu^\star$ on pooled $\hat\Sigma$
 & \textbf{per-class} $\mu_c^\star$ on $\Lambda_c$
 & per-block QP, independent across blocks
 & \textbf{single global} $\mu^\star$ on $\{\Lambda_c\}_{c=1}^{K}$ \\
Mixture-aware?
 & no
 & yes (per-class)
 & no (data-oblivious)
 & yes (cross-class, single multiplier) \\
R--D characterization
 & closed-form via reverse waterfilling
 & operational only
 & operational only
 & closed-form converse--achievability sandwich (Thms.~\ref{thm:water},\ref{thm:ach}) \\
Overhead beyond $R_{\rm cond}(D)$
 & $4$--$12$ dB mismatch
 & $H(C)/n$ + per-class slack
 & not analyzed (operational)
 & $H(C)/n$ only (Thm.~\ref{thm:ach}) \\
\bottomrule
\end{tabularx}
\end{table}

\subsection{Why a single global water level is information-theoretically optimal}

Let $\{(\pi_c,R_c,U_c,\Lambda_c=\diag(\lambda_{c,1},\ldots,\lambda_{c,n}))\}_{c=1}^{K}$
be the true Gaussian-mixture parameters. Conditioning on the label
$C$, the genie-aided per-dim rate--distortion characterization is a
\emph{joint} convex program in the per-mode distortions
$\{d_{c,i}\ge 0\}$:
\begin{align}
\label{eq:joint_rd_program}
\min_{\{d_{c,i}\}}\;\frac{1}{n}\sum_{c=1}^{K}\pi_c\sum_{i=1}^{n} d_{c,i}
\quad\text{s.t.}\quad
\frac{1}{n}\sum_{c=1}^{K}\pi_c\sum_{i=1}^{n}\frac12
\bigl[\log_2\!\tfrac{\lambda_{c,i}}{d_{c,i}}\bigr]_+ \le R.
\end{align}

\paragraph{(A) PrismQuant: one constraint, one Lagrange multiplier.}
Forming the Lagrangian with multiplier $\nu\ge 0$,
\[
\mathcal L(\{d_{c,i}\},\nu) =
\frac{1}{n}\sum_{c,i}\pi_c\!\left[d_{c,i} + \nu\cdot
\tfrac12\bigl[\log_2\!\tfrac{\lambda_{c,i}}{d_{c,i}}\bigr]_+\right] - \nu R,
\]
the KKT stationarity for any active mode is
\[
\pi_c \!\left[1 - \frac{\nu}{2 d_{c,i}\ln 2}\right] = 0
\quad\Longrightarrow\quad
d_{c,i}^\star \;=\; \frac{\nu}{2\ln 2} \;\triangleq\; \mu^\star
\quad\text{for all}\;(c,i),
\]
and the active set is $\mathcal A_c(\mu^\star) = \{i:\lambda_{c,i}>\mu^\star\}$.
The prior $\pi_c$ \emph{cancels} from both sides of the
stationarity condition: every extra bit on mode $(c,i)$ is paid and
rewarded with the same probability $\pi_c$. The rate constraint is the
\emph{only} coupling between modes, hence a \emph{single} multiplier
$\nu$, hence a \emph{single} global water level $\mu^\star$
(Theorem~\ref{thm:water}).

\paragraph{(B) WUTC: $K$ extra constraints, $K$ extra multipliers.}
WUTC \citep{EffrosChou95} uses a per-class bit budget $R_q$ for every
class: the bits spent inside class $c$ cannot be shared with any other
class. Equivalently, WUTC adds $K-1$ \emph{extra} equality constraints
to \eqref{eq:joint_rd_program},
\[
\frac{1}{n}\sum_{i=1}^{n}\frac12\bigl[\log_2\!\tfrac{\lambda_{c,i}}{d_{c,i}}\bigr]_+ \;=\; R_q,
\qquad c=1,\ldots,K,
\]
so the Lagrangian acquires $K$ independent multipliers
$\{\nu_c\}_{c=1}^{K}$ and the KKT condition becomes
$d_{c,i}^{\rm WUTC} = \nu_c/(2\ln 2)\triangleq \mu_c^\star$, a
\emph{per-class} water level that ignores the eigenvalue spectrum of
every other class.

\paragraph{(C) Strict dominance via constraint-set inclusion.}
The feasible set of WUTC is a proper subset of that of
\textsc{PrismQuant} (more constraints $\Rightarrow$ smaller feasible
set), so by elementary convex analysis,
\[
D_{\rm PQ}(R) \;\le\; D_{\rm WUTC}(R)
\qquad\text{for every } R\ge 0,
\]
with equality \emph{iff} the per-class allocation produced by WUTC
happens to satisfy $\mu_1^\star = \cdots = \mu_K^\star = \mu^\star$,
which occurs only in the degenerate symmetric case
$\Lambda_1 = \cdots = \Lambda_K$ (all classes have identical
eigenvalue spectrum). For \emph{any} non-degenerate mixture the
inequality is strict. The gap is concentrated on classes whose
eigenvalues are an outlier of the joint distribution: bits stuck in a
narrow-spectrum class with $\lambda_{c,i}\le\mu^\star$ would have
been more useful in a wide-spectrum class, but WUTC cannot move them.

\paragraph{(D) Information-theoretic interpretation.}
The same statement reads as a converse argument:
\textsc{PrismQuant} attains the genie-aided RD function
$R_{\rm cond}(D)$ up to $H(C)/n$ (Theorem~\ref{thm:ach}), so by
construction it is sandwiched between the unconditional RD function
$R_X(D)$ and $R_X(D)+H(C)/n$. WUTC, by contrast, is not sandwiched:
its operating point lies strictly above $R_{\rm cond}(D)+H(C)/n$ at
every non-degenerate mixture. Single-cov TC has an even larger gap
since it does not even use the label, paying the additional
\emph{mixture-mismatch} cost
$\mathrm{KL}\!\bigl(\sum_c\pi_c\mathcal N(\mu_c,R_c)\;\big\|\;\mathcal N(\hat\mu,\hat\Sigma)\bigr)$.

\subsection{Adaptive Multiple Transform (AMT/MTS): the standardized counterpart}
\label{app:amt_detail}

The library-of-transforms principle also appears in the HEVC and VVC
video-coding standards under the name \emph{adaptive multiple
transform} (AMT), later renamed multiple transform selection (MTS).
The AMT concept was first proposed by \citet{HanSaxenaRose10ASTC,
HanSaxenaMelkoteRose12AMT} as a joint optimization of spatial
prediction and block transform; \citet{SaxenaFernandes13AMT} extended
it into a mode-dependent DCT/DST scheme, \citet{ZhaoEMT16}
consolidated the design as enhanced multiple transform, and
\citet{BrossVVC21} documents its current form in VVC. The library is a
small set of \emph{fixed trigonometric transforms} (DCT-II together
with DST-VII and DCT-VIII variants), the branch is chosen by an
encoder-side rate--distortion search and signaled with a few
context-coded bits, and the bit allocation is governed by a
per-block quantization parameter from the standard.

AMT thus sits next to WUTC in the design space: both transmit a
branch index and apply a per-branch transform plus scalar
quantization, but AMT trades content-adaptivity for standardization,
fast trigonometric implementations, and hardware friendliness. It
uses neither a learned library nor a probabilistic source model, and
its rate--distortion behavior is documented only operationally, with
no analogue of the strict-dominance argument of \S(C). \textsc{PrismQuant}
differs from AMT in the same three respects that distinguish it from
WUTC: \emph{(i)} the library is derived from a Gaussian-mixture source
model instead of being hand-designed; \emph{(ii)} the branch is chosen
by a Bayes-optimal MAP classifier instead of by an RD search; and
\emph{(iii)} the bit allocation is summarized by a single global water
level with a matching closed-form converse, instead of by per-block
quantization parameters without an information-theoretic gap
analysis.

\subsection{Empirical confirmation on a synthetic mixture}

\begin{figure}[h]
\centering
\includegraphics[width=0.9\textwidth]{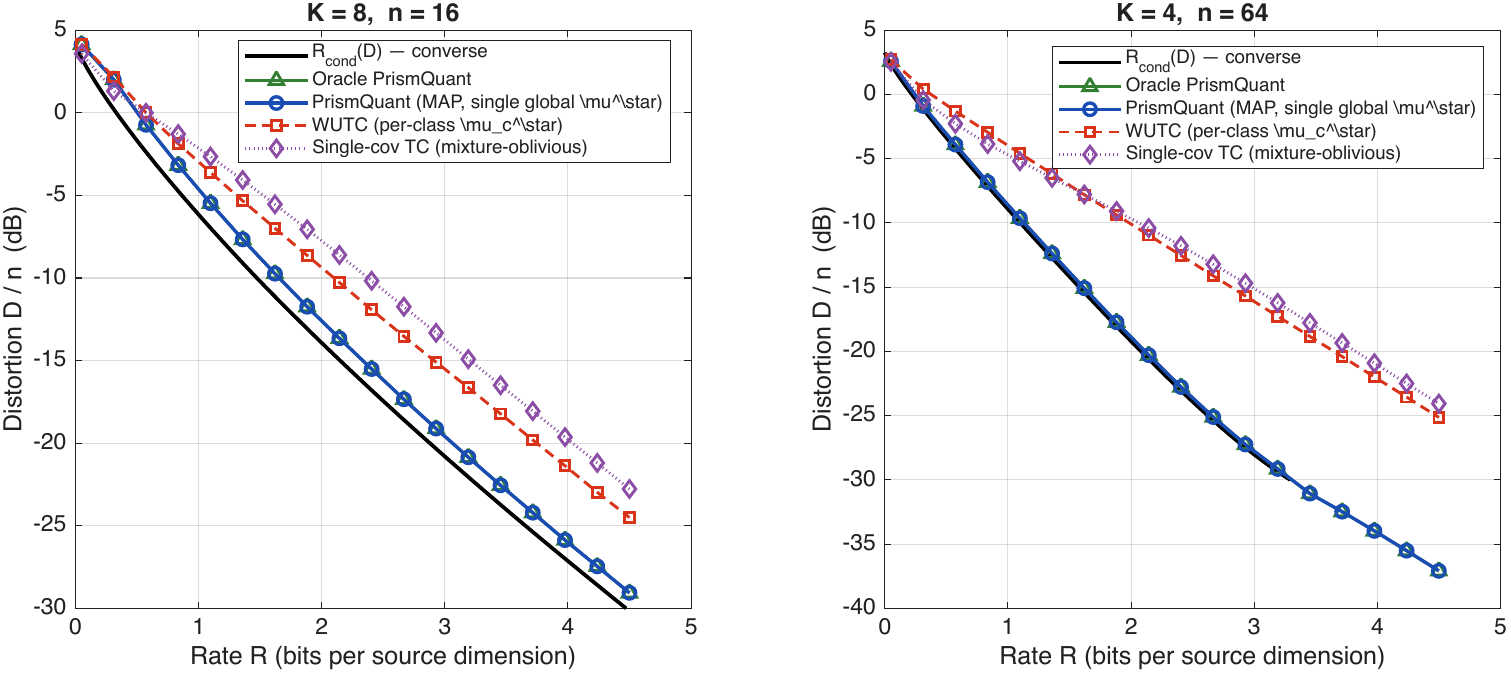}
\caption{Synthetic Gaussian-mixture RD comparison. \textsc{PrismQuant} sits
within the predicted $H(C)/n$ corridor of the converse;
WUTC pays the per-class allocation slack; single-covariance TC pays
the additional mixture-mismatch cost.}
\label{fig:tc_wutc_pq}
\end{figure}

Figure~\ref{fig:tc_wutc_pq} reports the operational RD curves of all
three baselines, together with the genie-aided converse $R_{\rm cond}(D)$
of Theorem~\ref{thm:converse}, on the synthetic mixture of
Section~\ref{sec:csi} ($K{=}8$, $n{=}16$, half wide-band and half
narrow-band classes; sample script
\texttt{matlab/exp\_synthetic\_pq\_vs\_hs.m}). At every measured rate
the empirical ordering is
\[
R_{\rm cond}(D)
\;\le\; \text{Oracle PQ}
\;\lesssim\; \text{PrismQuant (MAP)}
\;\le\; \text{WUTC}
\;\ll\; \text{Single-cov TC},
\]
with the gaps decomposing exactly as the analysis predicts:
PrismQuant--Converse $\approx$ $H(C)/n$ + ECSQ shaping
($\sim 0.255$ bit/active mode);
WUTC--PrismQuant $\approx$ per-class allocation slack (the inequality
proven in (C) above);
Single-cov TC--WUTC $\approx$ mixture-mismatch cost.

\subsection{Take-away}
The structural advantage of \textsc{PrismQuant} is
information-theoretic, not algorithmic. WUTC and \textsc{PrismQuant}
share the same encoder--decoder skeleton, but WUTC's per-class bit
budget is a \emph{strictly tighter} constraint than \textsc{PrismQuant}'s
single total-rate constraint, and tighter constraints produce weakly
larger optimal distortion. AMT/MTS faces the same structural gap from
the opposite direction: its transform library and quantizer are fixed
and standardized rather than data-driven, so it cannot match either
the per-class adaptivity of WUTC or the cross-class joint allocation
of \textsc{PrismQuant}. Single-covariance TC inherits both losses. The
only way to recover the slack is to merge all per-class budgets into a
single global one over a data-driven transform library---which is
exactly what \textsc{PrismQuant} does, and what
Theorem~\ref{thm:water} proves to be optimal.

\section{Additional Experiments} \label{app:exp_extra}

\subsection{Effect of $n$ on synthetic mixtures}

\begin{figure}[t]
    \centering
    \begin{subfigure}[t]{0.32\textwidth}
        \includegraphics[width=\textwidth]{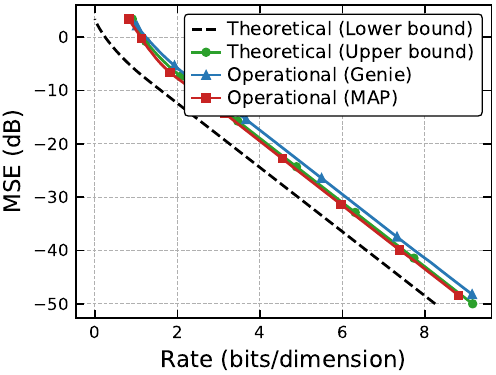}
        \caption{$n=4$}
    \end{subfigure}\hfill
    \begin{subfigure}[t]{0.32\textwidth}
        \includegraphics[width=\textwidth]{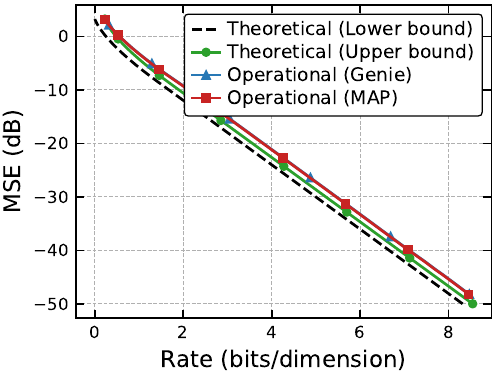}
        \caption{$n=16$}
    \end{subfigure}\hfill
    \begin{subfigure}[t]{0.32\textwidth}
        \includegraphics[width=\textwidth]{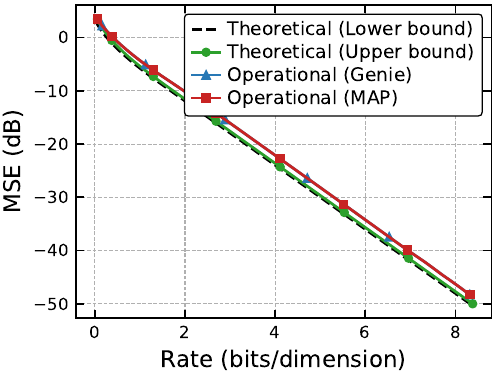}
        \caption{$n=64$}
    \end{subfigure}
    \caption{Effect of source dimension $n$ at fixed mixture order $K=16$. As $n$ increases, the gap between the theoretical lower and upper bounds progressively shrinks, consistent with Corollary~\ref{cor:sandwich}. Operational (Genie) closely tracks the theoretical upper bound, while Operational (MAP) even outperforms Operational (Genie) scheme when $n=4$.}
    \label{fig:synth_n}
\end{figure}

Figure~\ref{fig:synth_n} shows the RD performance of \textsc{PrismQuant} on synthetic Gaussian-mixture sources with fixed mixture order $K=16$ and varying source dimension $n\in\{4,16,64\}$. As the source dimension increases, the gap between the theoretical lower and upper bounds progressively shrinks, consistent with Corollary~\ref{cor:sandwich}, since the amortized component-label cost decreases with $n$. Moreover, both the genie-aided and MAP-based implementations closely approach the theoretical upper bound, demonstrating that practical \textsc{PrismQuant} nearly attains the predicted RD performance. Interestingly, in the low-dimensional regime ($n=4$), the MAP-based implementation slightly outperforms the genie-aided counterpart. This occurs because, when the mixture components are heavily overlapped relative to the source dimension, applying a component-matched KLT using a nearby estimated component can occasionally yield a more favorable RD trade-off than conditioning on the true generating component.

\subsection{Effect of $n$ on DeepMIMO}
\label{app:effect_of_n}

\begin{figure}[t]
    \centering
    \begin{subfigure}[t]{0.48\textwidth}
        \includegraphics[width=\textwidth]{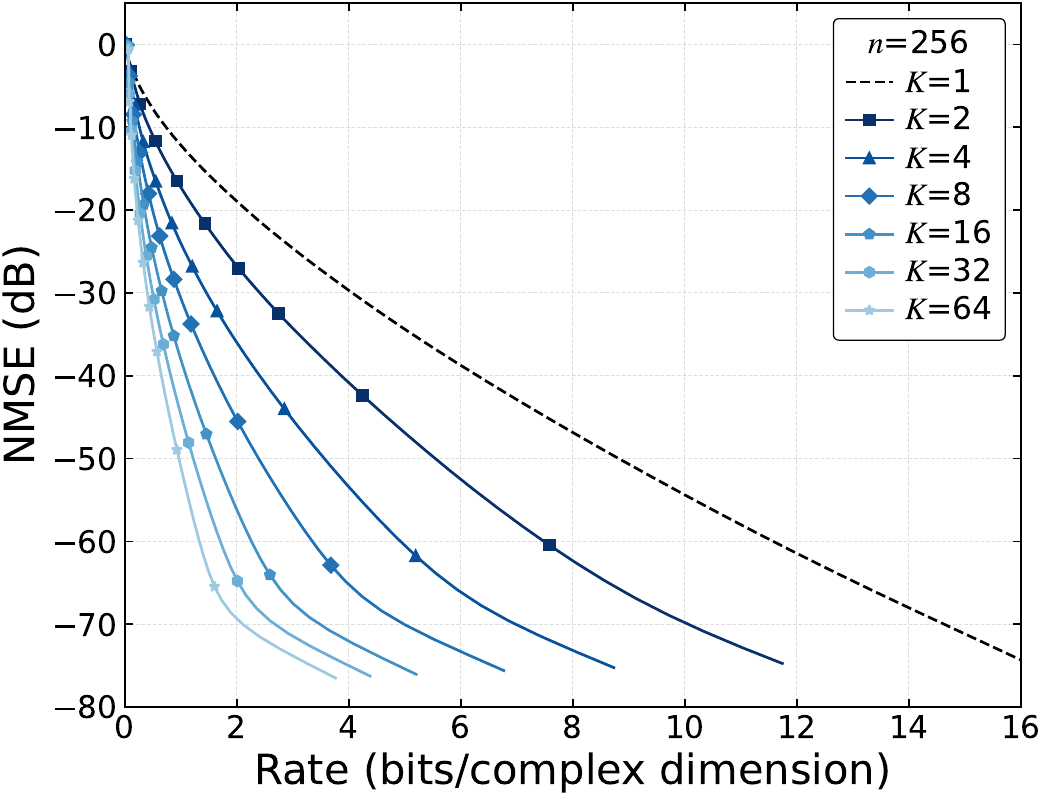}
        \caption{$n=256$}
    \end{subfigure}\hfill
    \begin{subfigure}[t]{0.48\textwidth}
        \includegraphics[width=\textwidth]{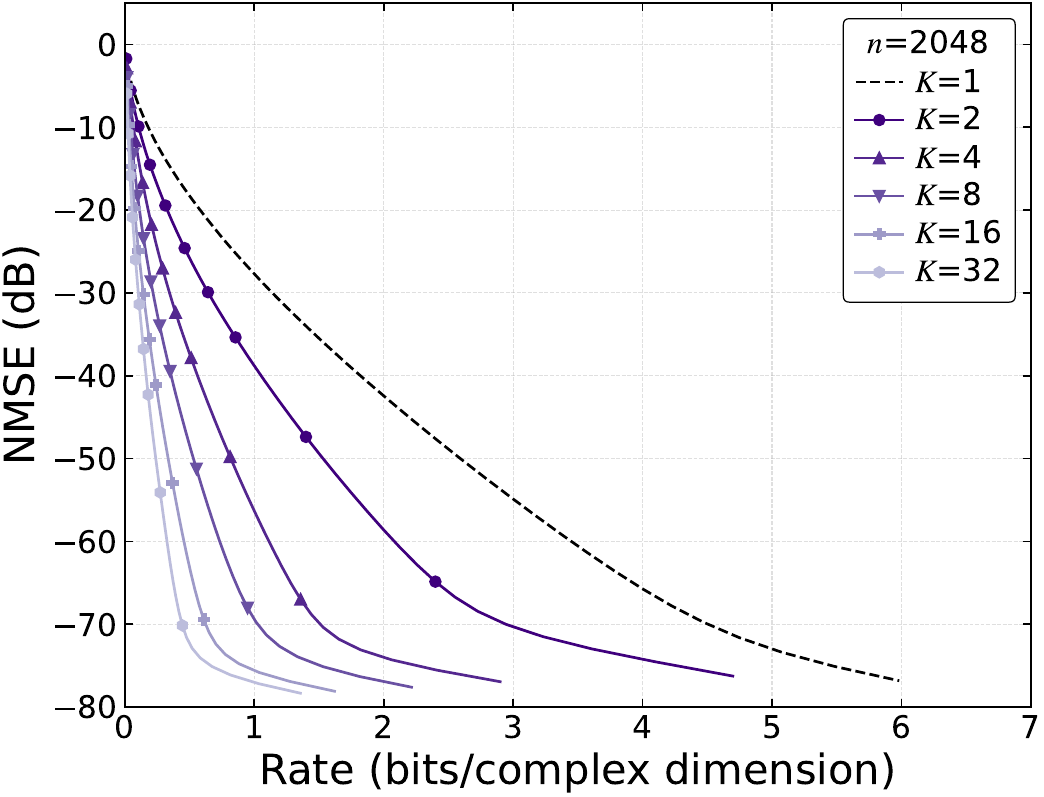}
        \caption{$n=2048$}
    \end{subfigure}
    \caption{Effect of block length $n$ on DeepMIMO at varying mixture
    order $K$. Together with Figure~\ref{fig:DeepMIMO} (b) ($n=1024$),
    these results show that for any fixed $K$, increasing $n$
    consistently improves RD performance across the entire rate range.}
    \label{fig:n_sweep}
\end{figure}

Figures~\ref{fig:DeepMIMO} (b) and \ref{fig:n_sweep} show the RD performance of \textsc{PrismQuant}
on the DeepMIMO dataset with fixed mixture order $K$ and varying block
length $n\in \{256, 1024, 2048\}$. Partitioning the channel into blocks of size $n$ implicitly assumes that distinct blocks are statistically independent;
since CSI is in fact strongly correlated across dimensions, smaller
$n$ introduces a mismatch with the true distribution, while larger
$n$ allows the mixture to capture longer-range dependencies and
yields a tighter approximation of the underlying source. As a result,
for any fixed mixture order $K$, increasing $n$ consistently shifts
the RD curve downward across the entire rate range, demonstrating
that more faithful distributional modeling directly translates into
improved compression performance.

\end{document}